\documentclass[reprint,preprintnumbers,nofootinbib,superscriptaddress,amsmath,amssymb,aps,prd]{revtex4-2}
\pdfoutput=1
\usepackage{graphicx}
\usepackage{physics}
\usepackage{orcidlink}
\usepackage{algorithm2e}
\usepackage{amsmath,amssymb}
\usepackage{dsfont}
\usepackage{multirow}
\usepackage{lineno}
\usepackage{siunitx}
\usepackage{tikz}
\usetikzlibrary{quantikz2}
\usepackage{bm}
\usepackage{braket}
\usepackage{graphicx}
\usepackage{subfigure}
\usepackage{caption}
\usepackage{ragged2e}
\usepackage{subcaption}
\graphicspath{{figs/}}
\usepackage[colorlinks=true,
            linkcolor=blue,      
            citecolor=olive,     
            filecolor=magenta,   
            urlcolor=blue]{hyperref}

\hypersetup{
    unicode=true,
    pdftoolbar=true,
    pdfmenubar=true,
    pdffitwindow=false,
    pdfstartview={FitH},
    pdftitle={spin-boson},
    pdfauthor={...}, 
    pdfsubject={spin-boson},
    pdfcreator={...},
    pdfproducer={}, 
    pdfkeywords={},
    pdfnewwindow=true,
    colorlinks=true,
    linkcolor=blue,
    citecolor=blue,
    filecolor=magenta,
    urlcolor=blue
}

\begin{document}

\title{Continuous-variable ADAPT-VQE for bosonic lattice models}

\author{Dimitrios Athanasakos}%
\thanks{Equal contribution, \href{mailto:dimitrios.athanasakos@hsbc.com.sg}{dimitrios.athanasakos@hsbc.com.sg}}

\affiliation{Department of Physics and Astronomy, Stony Brook University, Stony Brook, NY 11794, USA}
\affiliation{Quantum Technologies Group, HSBC, Singapore, 117439, Singapore}

\author{Gloria Tejedor-García}
\thanks{Equal contribution, \href{mailto:dimitrios.athanasakos@hsbc.com.sg}{gloria.tejedorgarcia@stonybrook.edu}}
\affiliation{Department of Physics and Astronomy, Stony Brook University, Stony Brook, NY 11794, USA}

\author{Jack Y. Araz}
\affiliation{Department of Physics and Astronomy, University College London, London, WC1E 6B, UK}
\affiliation{Department of Engineering, City St. George's, University of London, London, EC1V 0HB, UK}

\author{Mafalda Ram\^oa}
\affiliation{Department of Physics, Virginia Tech, Blacksburg, VA 24061, USA}
\affiliation{Virginia Tech Center for Quantum Information Science and Engineering, Blacksburg, VA 24061, USA}

\author{Bharath Sambasivam}
\affiliation{Department of Physics, Virginia Tech, Blacksburg, VA 24061, USA}
\affiliation{Virginia Tech Center for Quantum Information Science and Engineering, Blacksburg, VA 24061, USA}

\author{Sophia E. Economou}
\affiliation{Department of Physics, Virginia Tech, Blacksburg, VA 24061, USA}
\affiliation{Virginia Tech Center for Quantum Information Science and Engineering, Blacksburg, VA 24061, USA}

\author{Felix Ringer}
\affiliation{Department of Physics and Astronomy, Stony Brook University, Stony Brook, NY 11794, USA}

\begin{abstract}
We present a continuous-variable adaptive variational quantum eigensolver (CV-ADAPT-VQE). As concrete examples, we consider the ground-state preparation for (i) the Bose-Hubbard model and (ii) the bosonic Kitaev chain, including its extension with an on-site Kerr interaction. The former conserves the total boson number, while the latter conserves global parity. We construct symmetry-preserving operator pools tailored to each case and show, using GPU-based classical simulations, that CV-ADAPT-VQE results in significantly shallower circuits compared to Hamiltonian-based VQE approaches. Our results point toward direct applications in quantum simulations of condensed-matter systems, quantum chemistry, and high-energy physics.
\end{abstract}

\maketitle
\tableofcontents

\section{Introduction~\label{sec:intro}}

Quantum computing has witnessed rapid progress across a wide range of hardware platforms, including superconducting circuits, trapped ions, neutral atoms, and photonic systems~\cite{slussarenko_photonic_2019, kjaergaard_superconducting_2020, bruzewicz_trapped-ion_2019, browaeys_many-body_2020, gross_quantum_2017}. While much of this progress has focused on discrete-variable (DV) architectures based on qubits, various problems in quantum simulation and many-body physics are naturally formulated in high-dimensional or infinite-dimensional Hilbert spaces~\cite{RICO2018466, Bullock_2005, 10313758}. Encoding such systems into qubit registers requires truncations and encodings that can introduce significant computational overhead.

These considerations have motivated increasing interest in continuous-variable (CV) quantum computing, where the fundamental unit of information is a \emph{qumode}, i.e.\ a bosonic degree of freedom or harmonic oscillator described by canonical quadrature operators $\hat{q}$ and $\hat{p}$~\cite{PhysRevLett.82.1784, Menicucci_2006, Gu_2009, Weedbrook_2012}. Qumodes provide access to a formally infinite-dimensional Hilbert space through the Fock basis and therefore offer a natural framework for simulating bosonic systems, quantum fields, and lattice gauge theories. While CV quantum computing has often been associated with photonic platforms~\cite{tasca_continuous-variable_2011, bartlett_universal_2002, menicucci_one-way_2008}, recent advances have demonstrated high-fidelity control of bosonic modes across multiple leading quantum platforms, including superconducting microwave cavities~\cite{busnaina_native_2025, heeres_implementing_2017, wang_efficient_2020, blais_circuit_2021}, trapped-ion motional modes~\cite{fluhmann_encoding_2019, leibfried_quantum_2003, matsos_robust_2024, sutherland_universal_2021, lau_proposal_2012}, and ultracold atomic systems~\cite{bouchoule_neutral_1999, bohnmann_bosonic_2025, shaw_erasure-cooling_2025, kendell_deterministic_2024}. These developments establish qumodes as a versatile resource for quantum computations.

A central theoretical concept in CV quantum computing is universality. In the formulation by Lloyd and Braunstein, universal quantum computation can be achieved using Gaussian operations supplemented by a single non-Gaussian element, enabling the construction of arbitrary unitary transformations generated by finite polynomials in $\hat{q}$ and $\hat{p}$~\cite{PhysRevLett.82.1784}. This construction relies on commutator identities and Trotterization to systematically build higher-order polynomial Hamiltonians. More recently, this has been extended to hybrid settings involving both discrete and continuous variables, where universal gate sets can be constructed from combinations of qubit Pauli operators and polynomial functions of bosonic quadratures~\cite{Liu:2024mbr, araz_toward_2025}. These hybrid approaches provide a powerful route toward simulating interacting quantum systems that involve both spin and bosonic degrees of freedom. In addition, non-polynomial operators such as trigonometric gates can be realized in hybrid qubit-qumode architectures~\cite{Chalermpusitarak:2025cod,Rainaldi:2025ymn,McGarry:2026mkk}. In parallel, classical simulation methods for bosonic circuits have also been developed. For example, a recent coherent-state propagation framework provides simulation guarantees for circuits with only logarithmically many Kerr gates \cite{guseynov2026coherentstatepropagationcomputationalframework}. Together, these recent developments highlight the broad opportunities of CV and DV-CV hybrid quantum computing and motivate the exploration of algorithms that can fully exploit these capabilities.

These advances are particularly relevant for quantum simulations of strongly correlated bosonic systems and quantum field theories. Continuous-variable and hybrid qubit-qumode approaches have recently been explored for simulating lattice gauge theories and related models~\cite{Bauer:2022hpo,Marshall:2015mna,davoudi_toward_2021,Saner:2025nrq,Briceno:2023xcm,Ale:2024uxf,Crane:2024tlj,Abel:2024kuv,Ale:2025sxz,Abel:2025zxb,Gupta:2025xti}. More broadly, bosonic quantum simulation plays a central role in condensed-matter physics, quantum chemistry, and quantum optics, where collective excitations and field modes naturally arise. Efficient preparation of ground states and low-energy excitations in such systems remains a key challenge.

Variational quantum algorithms have emerged as a leading approach for ground-state preparation on near-term quantum hardware~\cite{Peruzzo_2014, Kandala_2017, Tilly_2022, Cerezo_2022, Yuan2019theoryofvariational, farhi2014quantumapproximateoptimizationalgorithm}. Among these, the Variational Quantum Eigensolver (VQE) has been widely adopted for approximating low-energy states of many-body quantum systems. However, its performance depends critically on the choice of ansatz. Fixed, hardware-efficient ansätze often suffer from barren plateaus or limited expressibility~\cite{McClean_2018, Ragone_2024, Fontana_2024, larocca2025barrenplateausvariationalquantum}, while physically motivated ansätze can require circuit depths that exceed the capabilities of current quantum devices and can be susceptible to local traps. The ADAPT-VQE algorithm addresses these limitations by constructing the ansatz dynamically~\cite{Grimsley:2018wnd, grimsley_adapt-vqe_2023, Tang_2021, PhysRevResearch.6.013254, Shkolnikov2023avoidingsymmetry, 10313885, 10.1063/5.0054822, Grimsley2023, Farrell:2023fgd, gustafson_surrogate_2024, Ramoa2025,Sambasivam:2025nbg}. Starting from a reference state, the algorithm iteratively selects operators from a predefined pool based on their energy gradients, tailoring the circuit to the structure of the target Hamiltonian. This adaptive procedure can significantly reduce circuit depth while achieving high accuracy.

\begin{figure*}[t]
    \centering
    \includegraphics[width=1\textwidth]{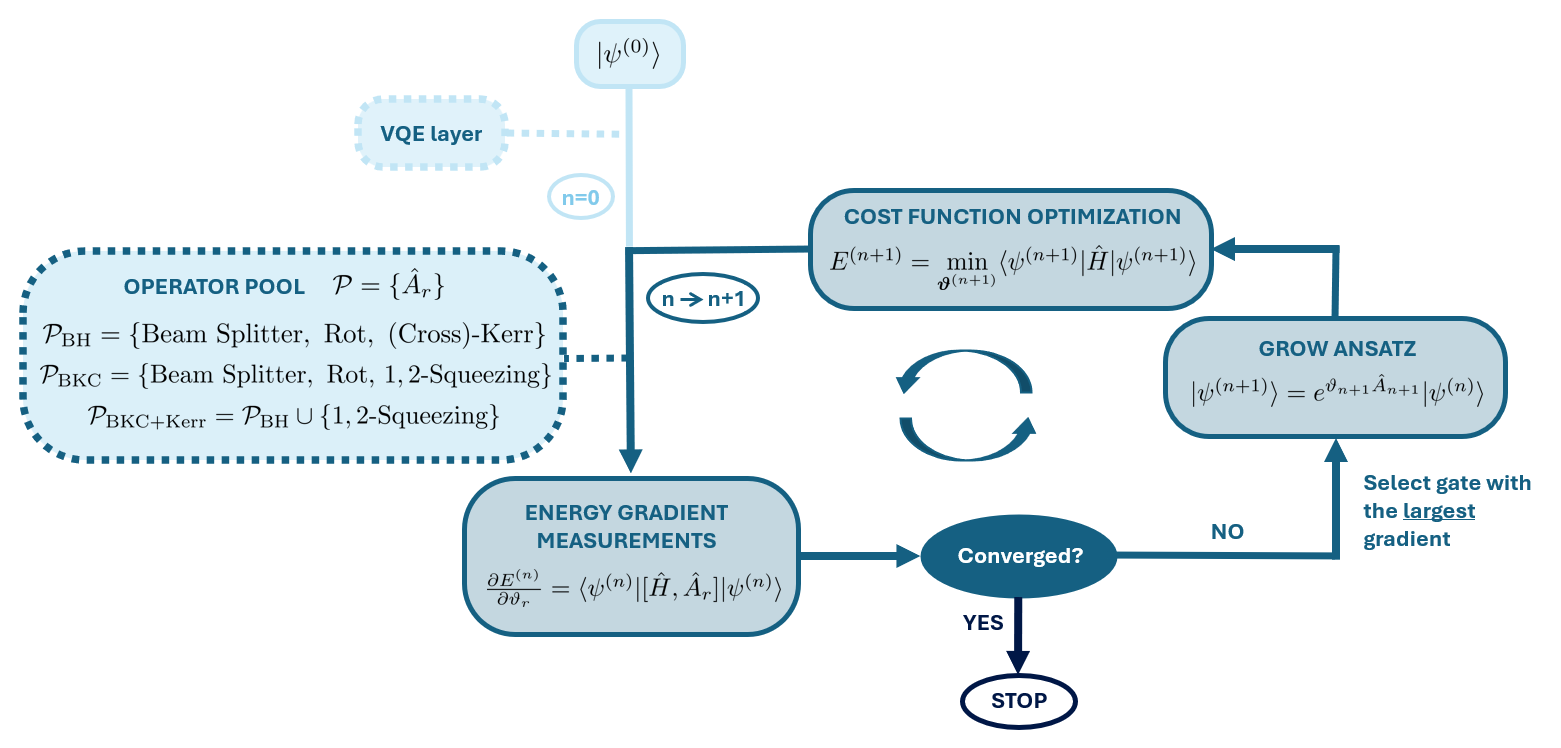}
    \caption{\justifying Schematic illustration of the CV-ADAPT-VQE ground state preparation algorithm presented in this work. The state after $n$ iterations is denoted by $\ket{\psi^{(n)}}$, which depends on the variational parameters of $n$ gates, collectively denoted by $\boldsymbol{\vartheta}^{\,(n)}$. Depending on the lattice Hamiltonian, we choose a suitable operator pool that incorporates symmetry constraints of the ground state.}
    \label{fig:adapt_diag}
\end{figure*}

In this work, we present a continuous-variable adaptive variational quantum eigensolver (CV-ADAPT-VQE). For comparison, we also consider Hamiltonian-based variational approaches in the continuous-variable setting, which we refer to as HVA-CV-VQE. Our approach constructs symmetry-preserving operator pools tailored to the structure of bosonic Hamiltonians, including Gaussian operations such as beam splitters, squeezing, and rotations, as well as non-Gaussian Kerr-type interactions. See Fig.~\ref{fig:adapt_diag} for a schematic illustration of the algorithm, which will be discussed in more detail below. We apply this framework to two representative lattice models: the Bose–Hubbard (BH) model and the bosonic Kitaev chain (BKC), including its extension with an on-site Kerr interaction. The BH model conserves the total boson number, while the BKC conserves global parity, allowing us to design operator pools that explicitly respect the relevant symmetries in each case. Continuous-variable variational algorithms have been explored in a related context. First HVA-CV-VQE results for the BH model were presented in Ref.~\cite{Yalouz:2021oyo}. More generally, Hamiltonian-based CV-VQE approaches for bosonic systems have been studied in Refs.~\cite{Kang:2023xfb,Dutta:2024zep,Dutta:2025kce,Chiari:2025lwq,dutta_solving_2026}. In addition, an extension of ADAPT-VQE for molecular vibrational structure was developed in Ref.~\cite{majland_vibrational_2025}. There, each vibrational mode is represented in a second-quantized basis mapped onto qubits through a direct encoding. The resulting creation and annihilation operators are explicitly not harmonic-oscillator ladder operators, and the circuit cost is measured in CNOT gates. Our approach instead acts natively on qumodes, expressing the adaptive ansatz in terms of bosonic ladder operators and native Gaussian and Kerr gates, and specifically targets bosonic lattice models. These developments provide important benchmarks and motivate the present work, where we systematically explore the use of CV-ADAPT-VQE for bosonic lattice models.

Using GPU-based classical simulations, we demonstrate that CV-ADAPT-VQE achieves significantly improved performance compared to Hamiltonian-based CV-VQE approaches, reaching high accuracy with substantially reduced circuit depth. These results highlight the potential of adaptive, symmetry-preserving variational algorithms for continuous-variable quantum computing and point toward direct applications in quantum simulations of condensed-matter systems, quantum chemistry, and high-energy physics. In particular, the combination of CV hardware capabilities, extended notions of universality, and adaptive algorithm design provides a promising pathway toward efficient quantum simulations of complex bosonic and field-theoretic systems.

The remainder of this paper is organized as follows. In section~\ref{sec:CV}, we review relevant aspects of continuous-variable quantum simulations. In section~\ref{sec:models}, we introduce the bosonic lattice models discussed in this work. We describe the HVA-CV-VQE and CV-ADAPT-VQE ground state preparation algorithms in section~\ref{sec:algorithms}, and present numerical results using classical simulations in section~\ref{sec:numerics}. We conclude and present an outlook in section~\ref{sec:conclusions}.

\section{Continuous-variable quantum simulations~\label{sec:CV}}

We first introduce the notation relevant to continuous-variable quantum computing and then review the elementary gate operations used throughout this work.

\subsection{Qumode representations}

The fundamental unit of information in continuous-variable quantum computing is the qumode. It corresponds to an ideal quantum mechanical harmonic oscillator with an associated infinite-dimensional Hilbert space $\mathcal{H}$. The state of a qumode is described by the quadrature operators $\hat q$ (position) and $\hat p$ (momentum), which satisfy the canonical commutation relation
\begin{equation}
    [\hat{q}, \hat{p}] = i \mathbb{I} \,.
\end{equation}
The eigenvalues of these operators are continuous variables. The operators are related to the bosonic creation and annihilation operators, satisfying $[\hat{a}, \hat{a}^\dagger] = \mathbb{I}$, via the linear transformations
\begin{equation}
    \hat{q} = \frac{\hat{a}^\dagger + \hat{a}}{\sqrt{2}}\,, \qquad \hat{p} = \frac{i(\hat{a}^\dagger - \hat{a})}{\sqrt{2}}\,.
\end{equation}
For the numerical simulation of interacting bosonic Hamiltonians, it is natural to work in the Fock representation. This basis $\{\lvert n\rangle\}_{n=0}^\infty$ is formed by the eigenstates of the particle number operator $\hat{N} = \hat{a}^\dagger \hat{a}$, such that $\hat{N}\lvert n\rangle = n \lvert n\rangle$. The action of the ladder operators in this basis is given by
\begin{equation}
    \hat{a} \lvert n\rangle = \sqrt{n}\,\lvert n-1\rangle\,, \qquad \hat{a}^\dagger \lvert n\rangle = \sqrt{n+1}\,\lvert n+1\rangle\,,
\end{equation} 
with $\lvert 0\rangle$ denoting the vacuum state. An arbitrary pure state $\lvert \psi\rangle \in \mathcal{H}$ admits the expansion 
\begin{equation}
\lvert \psi\rangle = \sum_{n=0}^\infty c_n \lvert n\rangle\,,\;\; \text{with}\;\;  \sum_{n=0}^\infty |c_n|^2 = 1\,. 
\end{equation}
Since the full Hilbert space is infinite-dimensional, classical numerical implementations require a truncation to a finite-dimensional subspace $\mathcal{H}_D = \text{span}\{\lvert 0\rangle, \ldots, \lvert D-1\rangle\}$, defined by a cutoff dimension $D$ sufficiently large to capture the relevant physics.

While continuous-variable systems can also be described using the Wigner quasiprobability distribution \cite{Belitsky_2004, Hagiwara_2016}, a formalism that is particularly useful for Gaussian states and operations \cite{ferraro2005gaussianstatescontinuousvariable, Adesso_2014}, throughout this work, we only utilize the Fock representation. This is motivated by the discrete particle-number nature of the target Hamiltonians, such as the BH model, which are naturally expressed and analyzed in the Fock basis.

\subsection{Gate operations}
\label{subsec:gates}

Universal quantum computations and state preparation over continuous variables are achieved through a set of elementary unitary operations~\cite{PhysRevLett.82.1784}. The single-mode Gaussian operations include the phase rotation gate
\begin{equation}
{\rm R}(\theta) = \text{exp}(i\theta \hat N)\,,
\end{equation}
and the displacement gate 
\begin{equation}
{\rm D}(\theta,\phi)
= \exp\!\left[\theta\!\left( e^{i\phi}\hat a^{\dagger}- e^{-i\phi}\hat a\right )\right],
\end{equation}
which performs a displacement of the wavefunction, as well as the single-qumode squeezing gate
\begin{equation}
{\rm S}(\theta,\phi)
= \exp\!\left[\frac{\theta}{2}\!\left(e^{-i\phi}\hat a^2 - e^{i\phi}\hat a^{\dagger 2}\right)\right],
\end{equation}
which introduces correlations between the quadratures of a single qumode. Multi-mode entanglement is generated via two-qumode Gaussian operations. The beam splitter gate is defined as
\begin{equation}
{\rm BS}(\theta,\phi)
= \exp\!\left[\theta\!\left(e^{i\phi}\hat a_1 \hat a_2^\dagger
- e^{-i\phi}\hat a_1^\dagger \hat a_2\right)\right]\,,
\end{equation}
which is number-conserving. Conversely, the two-mode squeezing gate
\begin{equation}
{\rm S}_2(\theta,\phi)
= \exp\!\left[\theta\!\left(e^{i\phi}\hat a_1 \hat a_2
- e^{-i\phi}\hat a_1^\dagger \hat a_2^\dagger\right)\right]
\end{equation}
generates entanglement while preserving parity. We note that ${\rm S}_2$ can be decomposed into a sequence of beam splitters and single-mode squeezing gates, though it is treated here as a distinct primitive for ansatz construction.

To access the non-Gaussian sector of the Hilbert space, which is essential for universal quantum simulations, the continuous variable gate set needs to include at least one non-Gaussian operation. A number-preserving non-Gaussian gate is the Kerr interaction
\begin{equation}
{\rm K}(\theta) = \text{exp}(i\theta \hat N^2)\,,
\end{equation}
which applies a phase shift proportional to the square of the boson number. Similarly, the Cross-Kerr gate
\begin{equation}
{\rm CK}(\theta) = \text{exp}(i\theta \hat N_1 \hat N_2),
\end{equation}
induces a conditional phase shift depending on the boson numbers in two distinct modes. 

\section{Bosonic lattice models~\label{sec:models}}

In this section, we introduce the three bosonic lattice models that serve as benchmarks for the performance of the CV-ADAPT-VQE algorithm presented in this work. We begin with the number-preserving Bose-Hubbard model and then turn to the bosonic Kitaev chain, which violates number conservation while preserving global parity. We consider both the pure model and an extension that includes an additional on-site Kerr interaction term.

\subsection{Bose-Hubbard model}

The BH model is a (1+1)-dimensional bosonic lattice model. The Hamiltonian is given by
\begin{align}
    \hat H_{\rm BH}=&\, -J \sum_{r}(\hat{a}_r^{\dagger} \hat{a}_{r+1}+\text{h.c.})-\frac{U}{2} \sum_r \hat{N}_r(\hat{N}_r-1)\nonumber \\
    &\,-\mu \sum_r \hat{N}_r\,,
    \label{eq:BoseHubbard}
\end{align}
where we sum over $r=1,\ldots,N_S$ lattice sites. Here, $\hat a_r,\hat a_r^\dagger$ are the bosonic lowering and raising operators at site $r$ that satisfy $[\hat a_r, \hat a_s^{\dagger}]=\mathbb{I}\delta_{r s}$. The number operator at site $r$ is given by $\hat N_r=\hat a_r^\dagger \hat a_r$. The first term in Eq.~(\ref{eq:BoseHubbard}) represents the kinetic energy, parameterized by the hopping amplitude $J\in \mathbb{R}$, which allows bosons to move between nearest-neighbor sites and favors delocalization. The second term represents the on-site interaction energy $U\in \mathbb{R}$ arising from boson scattering off of each other when they occupy the same mode. For $U>0$ ($U<0$), this term corresponds to the attractive (repulsive) Bose-Hubbard model, favoring localization (delocalization).  The third term corresponds to the chemical potential $\mu\in \mathbb{R}$, which controls the total number of particles in an eigenstate. The hopping and the chemical potential terms are Gaussian operations, since they are quadratic in the ladder operators $\hat a_r,\hat a_r^\dagger$. In contrast, the on-site interaction is a quartic operator in the ladder operators and can be written in terms of a non-Gaussian Kerr interaction and a Gaussian rotation term. 

\begin{figure*}[t]
        \includegraphics[width=0.4\textwidth]{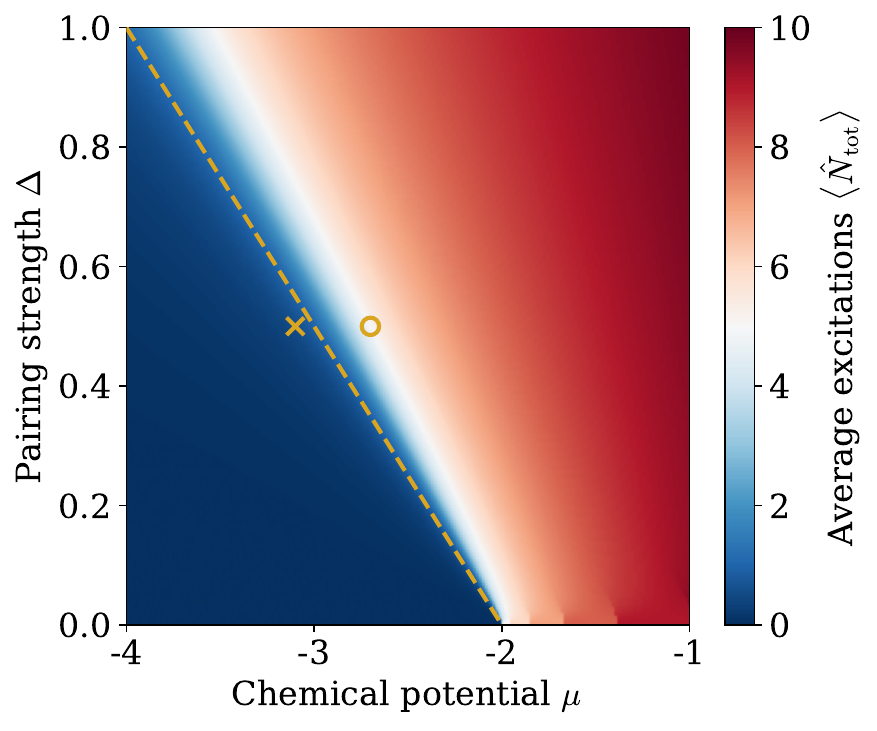}
        \hspace*{.7cm}
        \includegraphics[width=0.4\textwidth]{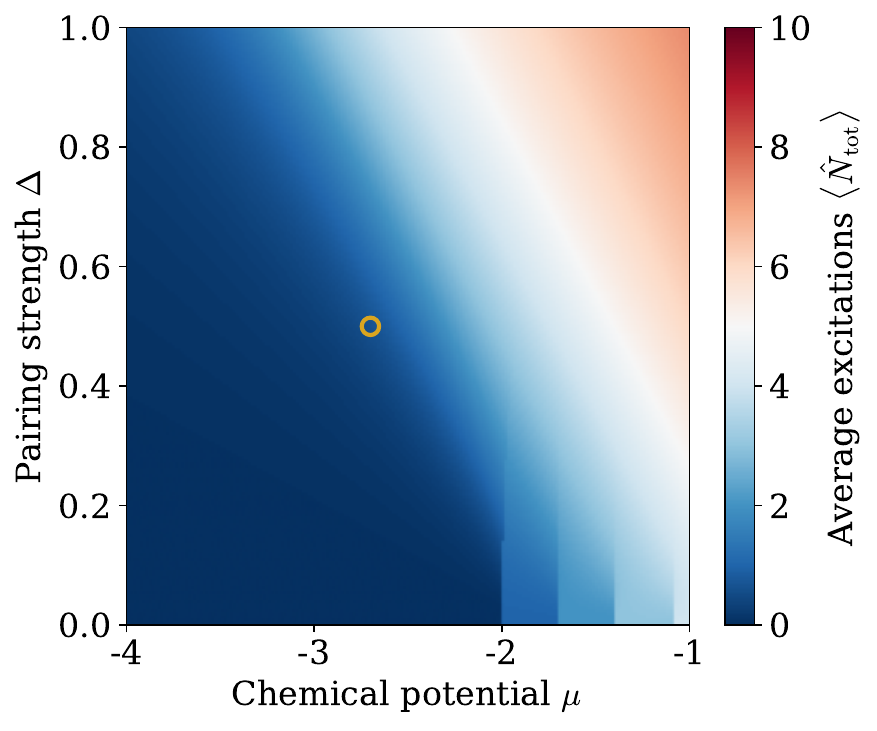}
    \caption{\justifying Expectation value of the total boson number across the lattice $\braket{\hat N_{\text{tot}}}$ for the ground state of the BKC model without (left) and with an additional on-site Kerr interaction (right). We show the result for different values of the model parameters $\Delta$ and $\mu$ for a lattice with $N_S = 3$ sites, $J=1$, $U=-1$ for the interaction term, periodic boundary conditions, and a local Hilbert space truncation of $D=6$. The dashed golden line in the left panel indicates the stability condition, while the markers 
    $\times$ and $\circ$ denote the parameter choices 
    $(\Delta=0.5,\mu=-3.1)$ and $(\Delta=0.5,\mu=-2.7)$, respectively, 
    which are used for the variational ground state preparation presented in section~\ref{sec:numerics}.}
    \label{fig:bkc_ntot}
\end{figure*}

The total number of excitations across the entire lattice is conserved since the total number operator
\begin{equation}
    \hat N_{\rm tot}=\sum _r \hat N_r \,,
\end{equation}
commutes with the Hamiltonian $[\hat N_{\rm tot},\hat H_{\rm BH}]=0$. This symmetry decomposes the BH Hamiltonian into disjoint sectors of fixed particle number since each eigenstate of the Hamiltonian must also be an eigenstate of the number operator $\hat N_{\rm tot}$. In this work, we will study only the problem of preparing the ground state for a fixed total number of excitations, rather than the global ground state, which can readily become gapless, making the problem intractable. The chemical potential term is diagonal in this subspace and corresponds to a constant energy drift $-\mu N_{\rm tot}$. Therefore, we set $\mu=0$ without loss of generality.

The competition between $J$ and $U$ gives rise to a zero-temperature phase transition in the global ground state. In the weakly interacting regime, $|J/U| \gg 1$, the kinetic term dominates, leading to a Bose-Einstein condensate delocalized over the entire lattice. In the strongly interacting regime, $|U/J| \gg 1$, the behavior depends on the sign of $U$. For repulsive interactions ($U<0$), particle-number fluctuations become energetically costly, giving rise to a Mott insulator phase \cite{fisher_boson_1989}. For attractive interactions ($U>0$), no Mott phase occurs; instead, particles tend to cluster on the same lattice site. This global behavior is also qualitatively reflected within each fixed $N_{\rm tot}$ sector of the Hamiltonian. For weak interactions, excitations are delocalized across all lattice sites. In the strongly repulsive regime, particles distribute as uniformly as possible to minimize double occupancy, whereas in the strongly attractive regime, they tend to cluster on the same lattice site.

To quantify the relative strength of the hopping term and the on-site interaction in Eq.~(\ref{eq:BoseHubbard}), we introduce the dimensionless parameter~\cite{Yalouz:2021oyo}
\begin{equation}
    \Lambda_{\rm BH}=\frac{N_{\text{tot}} U}{J}\,.
\end{equation}
The additional factor of $N_{\rm tot}$ in the numerator is included due to the relative scaling of the operators in Eq.~(\ref{eq:BoseHubbard}) with the boson number. Throughout this work, we consider the BH model with periodic boundary conditions (PBC) and truncate the qumode Hilbert space in the Fock basis at $D = N_{\text{tot}} + 1$ levels. For this particle-number-conserving model, the truncation of the local Hilbert space does not introduce numerical errors. Since the system is symmetric under a global rescaling of the parameters, we will set the hopping amplitude $J$ equal to 1 for the numerical simulations presented below. 

\subsection{Bosonic Kitaev chain}

Next, we consider the BKC model~\cite{wang_quantum_2022}. Similar to the BH model described above, the BKC is a (1+1)-dimensional lattice model. The analogous model with fermionic degrees of freedom is the fermionic Kitaev chain~\cite{Kitaev:2000nmw}. See also Refs.~\cite{McDonald:2018cdg,Busnaina:2023qvd}. Different versions of the BKC model appear in the literature, involving either single- or two-qumode squeezing terms. The ground state of the model with the single-qumode squeezing terms is generally expected to be less entangled and therefore easier to prepare. Here, we focus on the version with the two-qumode squeezing terms. The most general Hamiltonian of the BKC with two-qumode squeezing is given by
\begin{align}\label{eq:BosonicKitaevChain}
    \hat H_{\rm BKC}=&\, -J \sum_{r}(\hat{a}_r^{\dagger} \hat{a}_{r+1}+\text{h.c.})-\Delta  \sum_r  (\hat{a}_r \hat{a}_{r+1}+\text{h.c.}) \nonumber \\
    &\,-\mu \sum_r \hat{N}_r\,.
\end{align}
Here, $J, \Delta \in \mathbb{C}$ denote the hopping amplitude and pairing strength, respectively, and the sum runs over all $r=1,\dots,N_S$ lattice sites. While the general BKC model with complex pairings has non-trivial topological properties, we restrict our analysis here to the time-reversal symmetric limit where $J, \Delta \in \mathbb{R}$. This choice suppresses topological phenomena but retains the essential squeezing dynamics relevant to the ground state preparation algorithm explored in this work.

\begin{figure*}[t]
    \centering
    \scalebox{0.72}{
    \begin{quantikz}[row sep=0.5cm, column sep=0.15cm] 
        &\gate[2]{\text{BS}(\theta_1, \phi_1)}
        &
        & \permute{1,3,2}
        &\gate[2]{\mathrm{BS}(\theta_3,\phi_3)} 
        &\gate{{\rm K}(\theta_4)} & \gate{{\rm R}(\theta_7)} 
        &\\
        &
        & \gate[2]{\text{BS}(\theta_2, \phi_2)} 
        &
        &
        & \gate{{\rm K}(\theta_5)} 
        & \gate{{\rm R}(\theta_8)}  
        &\\
        &
        &
        &
        &
        & \gate{{\rm K}(\theta_6)} 
        & \gate{{\rm R}(\theta_9)} &\\
    \end{quantikz}
    \hspace*{0.2cm}
    \begin{quantikz}[row sep=0.5cm, column sep=0.15cm]
        &\gate[2]{\text{S}_2(\theta_1, \phi_1)}
        &
        & \permute{1,3,2}
        & \gate[2]{\mathrm{S}_2 (\theta_3,\phi_3)}
        & 
        & \gate[2]{{\rm BS}(\theta_5, \phi_5)}
        &\permute{1,3,2}
        &\gate[2]{\mathrm{BS} (\theta_6,\phi_6)}
        & \gate{{\rm R}(\theta_7)} 
        &  \\
        &
        & \gate[2]{\text{S}_2(\theta_2, \phi_2)} 
        &
        &
        & \gate[2]{{\rm BS}(\theta_4, \phi_4)}
        & 
        &
        &
        & \gate{{\rm R}(\theta_8)} 
        & \\
        &
        &
        &
        &
        &
        &
        &
        &
        & \gate{{\rm R}(\theta_9)} & \\
    \end{quantikz}
    }
    \caption{\justifying Hamiltonian-based VQE ans\"atze for the Bose-Hubbard model (left) and the bosonic Kitaev chain (right) for $N_S=3$ lattice sites and periodic boundary conditions. For the bosonic Kitaev chain with an additional on-site Kerr interaction, another set of corresponding single-qumode Kerr gates is included previous to the rotation gates.}
    \label{fig:HVA_CV_VQE}
\end{figure*}
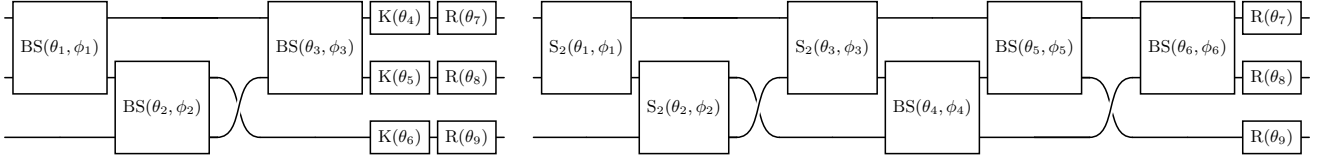

Two key features distinguish this model from the BH model. First, the pairing terms explicitly break the continuous $U(1)$ particle-number symmetry, $[\hat N_{\rm tot}, \hat H_{\rm BKC}] \neq 0$. However, the system retains a residual $\mathbb{Z}_2$ symmetry corresponding to the conservation of global boson-number parity, $\hat P = (-1)^{\hat N_{\rm tot}}$. The global ground state resides in the even-parity sector, a property that follows from its adiabatic continuity when starting from the vacuum state, which is the ground state of the non-interacting theory ($\Delta \to 0$). Due to conservation of global parity, the Hamiltonian is block diagonal in the two sectors. Analogous to the BH model described above, we limit ourselves to the ground state preparation in one of the two sectors. Second, the Hamiltonian is strictly quadratic in the bosonic operators. As a result, the ground state is Gaussian, and the problem is efficiently tractable classically. In particular, the spectrum can be obtained exactly via a Bogoliubov-de Gennes (BdG)~\cite{Bogolyubov:1958se, degennes1999} transformation. Despite its classical solvability, the BKC model is a nontrivial benchmark for variational ground-state preparation, allowing studies of circuit-depth requirements and finite-cutoff effects in the local qumode Hilbert space. In addition, it provides a natural starting point for non-Gaussian extensions, as discussed in more detail below. In the following, we fix the energy scale by setting the hopping amplitude to $J=1$.

In contrast to its fermionic counterpart, the pairing interaction in the BKC model can lead to an unbounded spectrum, depending on the choice of model parameters. For completeness, we provide a more detailed derivation of the stability conditions of the BKC model in Appendix~\ref{app1}. The resulting energy spectrum is given by
\begin{align}
E(k) &\, = \pm \sqrt{\xi_k^2 - \Delta_k^2}\,,\; \text{with} \nonumber\\
\xi_k &\, = -2J\cos k - \mu\,, \;\;
\Delta_k = 2\Delta\cos k \,,
\end{align}
where $k$ is the Fourier conjugate to the position variable. For $\xi_k > 0$, adding bosons increases the energy, and the system remains bounded. In contrast, for $\xi_k < 0$, adding bosons decreases the energy, leading to an instability. To ensure that the energy eigenvalues are real (i.e., to avoid a dynamical instability), the stability condition is
\begin{align}\label{eq:dispersion}
    \xi_k > |\Delta_k| \;\;\;\Rightarrow\;\;\; \mu < -2\,(J + \Delta)\,.
\end{align}
In Fig.~\ref{fig:bkc_ntot}, we show the ground-state expectation value of the total boson number operator $\braket{\hat N_{\rm tot}}$ for different values of the model parameters $\Delta$ and $\mu$ for a finite truncation $D$ of the local qumode Hilbert space. As expected, we observe that above the line $\Delta= -\mu/2-1$, the average number of bosons grows rapidly. Although the maximum allowed value for $N_{\text{tot}}$ is 15, corresponding to full occupancy of all sites, the maximum value achieved is $\braket{\hat N_{\text{tot}}} \approx 10$, which indicates that $\ket{\Psi_{0}} \propto \ket{D-1}^{\otimes N_S} + \epsilon \ket{\bar{\psi}}$  with $\ket{\bar{\psi}}$ corresponding to a non-maximum occupancy state. This factor $\epsilon$, which is not necessarily small, will make the optimization process much harder above the stability boundary $\Delta= -\mu/2-1$. The finite local Hilbert-space truncation acts as a hard ultraviolet (UV) regulator by limiting local bosonic occupation numbers, thereby bounding the spectrum and rendering the otherwise unstable region well-defined. This admits a direct analogy to quantum field theory, where UV regulators control high-energy modes and enable controlled studies near instability thresholds.

In addition, we consider the BKC model where an additional on-site Kerr interaction is included 
\begin{align}\label{eq:BosonicKitaevChain_Kerr}
\hat H_{\rm BKC+Kerr}= \hat H_{\rm BKC}-\frac{U}{2} \sum_r \hat{N}_r(\hat{N}_r-1)\,.
\end{align}
In the repulsive case ($U<0$), the on-site interaction can stabilize the otherwise unstable BKC model. The Kerr term can be viewed as a soft regularization of the BKC Hamiltonian, in contrast to the hard regularization imposed by truncating the local Hilbert space, as discussed above. While this extension continues to conserve global parity, the inclusion of the non-Gaussian Kerr term renders the model no longer purely Gaussian. This makes it a particularly relevant benchmark for assessing the performance of CV-ADAPT-VQE in more general, interacting continuous-variable systems. 

\begin{table*}[t]
\centering
\begin{tabular}{ll|ccc|ccc}
\hline\hline
 &  & \multicolumn{3}{c|}{\textbf{HVA-CV-VQE}} & \multicolumn{3}{c}{\textbf{CV-ADAPT-VQE}} \\ 
\textbf{Gate} &  & \textbf{BH} & \textbf{BKC} & \textbf{BKC+K} 
& \textbf{BH} & \textbf{BKC} & \textbf{BKC+K} \\ \hline
Beam Splitter & ${\rm BS}(\theta, \phi)$ 
& $\checkmark$ & $\checkmark$ & $\checkmark$
& $\checkmark$ & $\checkmark$ & $\checkmark$ \\

Phase Rotation  & ${\rm R}(\theta)$ 
& $\checkmark$ & $\checkmark$ & $\checkmark$
& $\checkmark$ & $\checkmark$ & $\checkmark$ \\

Kerr & ${\rm K}(\theta)$ 
& $\checkmark$ & -- & $\checkmark$
& $\checkmark$ & -- & $\checkmark$ \\

Cross-Kerr & ${\rm CK}(\theta)$ 
& -- & -- & --
& $\checkmark$ & -- & $\checkmark$ \\

Two-mode Squeezing & ${\rm S}_2(\theta, \phi)$ 
& -- & $\checkmark$ & $\checkmark$
& -- & $\checkmark$ & $\checkmark$ \\

Single-mode Squeezing & ${\rm S}(\theta, \phi)$ 
& -- & -- & --
& -- & $\checkmark$ & $\checkmark$ \\

\hline\hline
\end{tabular}
\caption{\justifying Overview of the gate set corresponding to the Hamiltonian Variational ansatz (HVA) and the operator pools for the CV-ADAPT-VQE used in this work. The HVA is restricted to unitary operations generated by the terms present in the system Hamiltonian. In contrast, the CV-ADAPT-VQE pools are augmented with additional generators that respect the global symmetries of the model. The different unitary gates are introduced in Section~\ref{subsec:gates}.}
\label{table:gates}
\end{table*}

\section{Variational ground state preparation~\label{sec:algorithms}}

In this section, we outline the variational algorithms employed to approximate the ground states of the BH and the BKC models. We begin by introducing the Hamiltonian Variational Ansatz (HVA) within the CV-VQE framework, which serves as a physically motivated benchmark, see also Ref.~\cite{Yalouz:2021oyo}. Next, we will discuss the continuous-variable extension of the ADAPT-VQE algorithm tailored to bosonic systems, which is the main focus of this work.

\subsection{HVA-CV-VQE}

To benchmark the performance of CV-ADAPT-VQE, we also consider a fixed-structure (static-ansatz) Hamiltonian Variational Ansatz (HVA)~\cite{Wecker2015, Ho:2019hyv, Cade:2020owo, Wiersema_2020}, which serves as a reference variational strategy for comparison. In the HVA approach, the variational circuit is constructed from the terms of the system Hamiltonian itself, rather than adaptively selecting operations during the optimization. For a Hamiltonian decomposed as $\hat{H} = \sum_k \hat{H}_k$, the ansatz is structured as a parameterized unitary evolution of depth $P$, where each layer $l$ applies the unitary exponentials of these components sequentially, mimicking a Trotterized time evolution:
\begin{equation}
    \hat{U}(\boldsymbol{\vartheta}) = \prod_{l=1}^P \left( \prod_k e^{-i \vartheta_{l,k} \hat{H}_k} \right) \,
    \label{eq:HVA_ansatz}
\end{equation}
Here, $\boldsymbol{\vartheta} = \{ \vartheta_{l,k} \}$ are the variational parameters that need to be optimized by minimizing the energy expectation value
\begin{equation}
     E(\boldsymbol{\vartheta}) = \langle \psi(\boldsymbol{\vartheta}) | \hat{H} | \psi(\boldsymbol{\vartheta}) \rangle\,.
\end{equation}
The physical motivation for this structure is based on adiabatic state preparation \cite{farhi2000, Santoro_2006}: in the limit of large circuit depth $P$, the ansatz is sufficiently expressive to approximate the adiabatic evolution from a trivial product state to the interacting ground state. However, the optimization of the parameters makes it a distinct unitary from the one that would be obtained by directly trotterizing the evolution operator generated by the adiabatic evolution.

For the models considered here, the circuit decomposition naturally follows the separation of the Hamiltonian into kinetic and interaction terms. For the BH model, the circuit layers alternate between two-qumode beam splitter operations associated with the kinetic sector and on-site rotations and non-Gaussian Kerr interactions. Note that, unlike the ansatz employed in Ref.~\cite{Yalouz:2021oyo}, our HVA layer includes rotation gates. In contrast, the BKC ansatz consists entirely of Gaussian operations, namely beam splitters, phase rotations, and squeezing gates. For the BKC model with the Kerr interaction, an additional set of Kerr gates is included. A central distinction among the three models is therefore the presence or absence of non-Gaussian terms. While the BKC ansatz explores only the manifold of Gaussian states, the inclusion of Kerr interactions generates non-Gaussianity and introduces quantum resources that are generally beyond those efficiently accessible by classical Gaussian simulation methods~\cite{Bartlett:2002geb}. However, different notions of complexity govern ground-state preparation. Non-Gaussian operations determine properties such as Wigner-function negativity or quantum magic, which are related to classical simulability, whereas the variational circuit depth required for state preparation is primarily controlled by physical properties of the Hamiltonian, such as its spectral gap. In particular, larger gaps generally permit shallower circuits, consistent with the adiabatic state-preparation picture that the variational ansatz and its adaptive extension aim to approximate.

A key feature of the HVA in both discrete and continuous-variable settings is its inherent symmetry-preserving character. Since the generators of the ansatz are derived directly from the Hamiltonian, they naturally preserve the symmetries of the problem. For the models considered here, either the total boson number (BH model) or global parity (BKC with and without Kerr interaction) is conserved. Earlier results suggest that such problem-inspired ans\"atze are less susceptible to the barren plateau problem than generic parameterized circuits, resulting in a more favorable optimization landscape for classical optimizers~\cite{Park_2024, McClean_2018, larocca2025barrenplateausvariationalquantum,arrasmith_2021}.

The circuits of the variational ansätze for the two models are shown in Fig.~\ref{fig:HVA_CV_VQE}, and the relevant gates are summarized in Table~\ref{table:gates}. In this work, we impose periodic boundary conditions for all models. Accordingly, the total number of variational gates per layer is $3\,N_S$ in both the BH and BKC models. When including the on-site interaction in the BKC model, an additional $N_S$ Kerr gates are required. For the BKC models, the vacuum $\ket{0}^{\otimes N_S}$ is chosen as the initial state, since it is the exact ground state of the $\Delta \to 0$ model in the stable regime. For the BH model, the initial state needs to balance competing requirements: it should be easy to prepare, and it should avoid strong localization, which suppresses gradients and can stall the CV-ADAPT-VQE convergence. In particular, since $\Lambda_{\rm BH}<0$ favors delocalization in the lattice, we choose an initial Fock state in which the $N_{\rm tot}$ bosons are spread over all sites. For $\Lambda_{\rm BH}>0$, where the attractive interaction favors clustering, we instead distribute the bosons over every other site. When $N_{\rm tot}$ is not divisible by the number of occupied sites, the remaining bosons are assigned to nearby occupied sites. These choices provide simple initial states that are not strongly localized.

\subsection{CV-ADAPT-VQE}

The main focus of this work is the extension of the ADAPT-VQE algorithm of Ref.~\cite{Grimsley:2018wnd} to continuous-variable quantum computing, denoted by CV-ADAPT-VQE. A schematic illustration of the algorithm for the bosonic lattice models considered here is shown in Fig.~\ref{fig:adapt_diag}, which we discuss in the following. The variational ansatz is constructed iteratively by adding unitary bosonic operators one at a time. The generators $\hat A_r$, which can be read off from the unitary operators in section~\ref{subsec:gates}, are selected from an operator pool, denoted by ${\cal P}$. For each model, the operator pool is constructed to respect the symmetries of the Hamiltonian. 

For the BH model, we include all number-preserving gates, while for the BKC model, we exclusively include Gaussian operations that preserve global parity. For both the BKC model with and without an on-site Kerr interaction, the operator pool includes beam splitters and two-mode squeezing gates acting beyond nearest-neighbor lattice sites, corresponding to all-to-all connectivity. For the BH model, however, we find that restricting the pool to nearest-neighbor operators improves performance. For the BKC model with a Kerr interaction, we additionally include non-Gaussian, parity-preserving operators such as Kerr and cross-Kerr gates. In Table~\ref{table:gates}, we summarize the unitary operators included in the three operator pools. In the two-parameter case, the unitary operators added to the ansatz are constructed as 
\begin{equation}
\hat O_r(\vartheta_r) = \text{exp}(\theta_r \hat A_r(\phi_r))\,,
\end{equation}
where $\vartheta_r=(\theta_r,\phi_r)$ denotes the variational parameters associated with the operator $\hat A_r \in \mathcal{P}$. For single-parameter gates, we have $\vartheta_r = \theta_r$  and
\begin{equation}
\hat O_r(\vartheta_r)=\text{exp}(\theta_r \hat A_r)\,.
\end{equation}
After sequentially appending $n$ unitary operators, the resulting state is denoted by $\ket{\psi^{(n)}}$. The state depends on the parameters of $n$ gates, which we collectively denote by $\boldsymbol{\vartheta}^{(n)}$. For a given ansatz, the algorithm minimizes the ground state energy $E^{(n)}(\boldsymbol{\vartheta}^{(n)})=\braket{\psi^{(n)}|\hat H|\psi^{(n)}}$.

At each iteration, the gradient associated with every operator in the pool is evaluated. For single-parameter gates, we have
\begin{equation}
    g_r = \left.\frac{\partial E^{(n)}}{\partial \theta_r}\right|_{\theta_r=0}
    = \langle \psi^{(n)} | [\hat H, \hat A_r] | \psi^{(n)} \rangle \,.
    \label{eq:grad}
\end{equation}
For two-parameter gates with $\vartheta_r=(\theta_r,\phi_r)$, the quantity used for gate selection is the Euclidean norm of the corresponding gradient vector
\begin{equation}
    \|\mathbf g_r\|_2
    = \left.\sqrt{ \left( \frac{\partial E^{(n)}}{\partial \theta_r} \right)^2
    +
    \left( \frac{\partial E^{(n)}}{\partial \phi_r} \right)^2
    }\right|_{\vartheta_r=0}\,.
\end{equation} 
For the two-parameter gates considered here, however, we find $\partial E^{(n)}/\partial \phi_r|_{\vartheta_r=0}=0$. Therefore, gate selection is determined entirely by the
$\theta_r$-dependence of $E^{(n)}$. Experimentally, these expectation values can be obtained using parameter-shift rules, consisting of different number-resolving or homodyne measurements~\cite{Wierichs:2021nwf}. The operator with the largest gradient magnitude is selected, and the ansatz is updated according to
\begin{equation}
|\psi^{(n+1)}\rangle
= \hat O_{n+1}(\vartheta_{n+1}) |\psi^{(n)}\rangle \,.
\end{equation}
For the subsequent iteration, the parameters are initialized recursively as
\begin{equation}
\boldsymbol{\vartheta}^{(n+1)}
=
\{\boldsymbol{\vartheta}^{(n)},\vartheta_{n+1}^{\rm init}\}\,,
\end{equation}
where $\vartheta_{n+1}^{\rm init}=0$ for single-parameter gates and
$\vartheta_{n+1}^{\rm init}=(0,0)$ for two-parameter gates.
The optimization is then repeated by minimizing the energy expectation value
$E(\boldsymbol{\vartheta}^{(n+1)})$.

For qubits, the ADAPT-VQE algorithm is typically terminated when the norm of the gradient vector, see Eq.~(\ref{eq:grad}), falls below a prescribed threshold, i.e., $|\mathbf{g}| < \epsilon$. Here, to enable a one-to-one comparison between HVA-CV-VQE and CV-ADAPT-VQE, we instead adopt a unified stopping criterion where each algorithm is terminated either after applying approximately 160 gates or once the energy difference with respect to the ground state, obtained using exact diagonalization, satisfies $\Delta E = |E-E_0| < \epsilon$, with $\epsilon = \mathcal{O}(10^{-10})$, which is close to machine precision.

A known limitation of greedy algorithms such as CV-ADAPT-VQE is their sensitivity to the initial state, which can lead to stagnation if the initial gradients vanish. While the vacuum state $\ket{0}^{\otimes N_S}$ is a suitable starting point for the BKC model, the ground states of the BH and interacting BKC models are inherently non-Gaussian. For any state $\ket{\psi}$ that is a product state in the Fock basis, the gradient $g_r$ in Eq.~\eqref{eq:grad} vanishes trivially for any gate whose generator is diagonal in that basis. This applies to both the Kerr and Cross-Kerr operators used here. To circumvent this, we employ a ``warm-start'' using a single layer of the HVA. This initialization provides a physically motivated prior that activates the non-Gaussian sector, and these HVA parameters are further refined during the subsequent optimization. We show in Appendix~\ref{app:warmup} that the final variational energy is not very sensitive to the specific form of this warm-up circuit.

\begin{figure*}[t]
    \centering
    \subfigure[]{
        \includegraphics[width=0.4\textwidth]{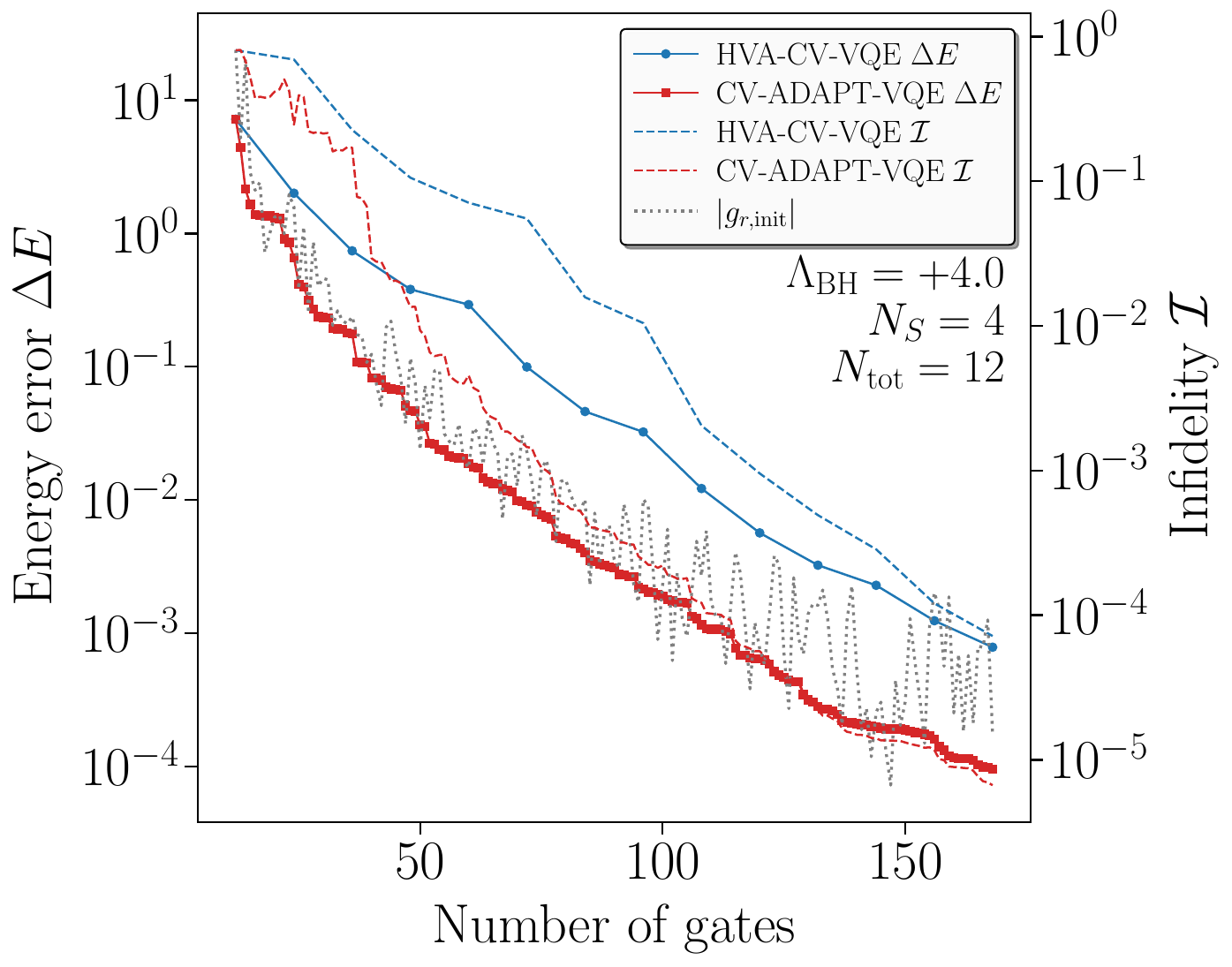}
        \label{fig:BH_ATRAC}
    }\hspace{0.8cm}  
        \subfigure[]{
        \includegraphics[width=0.4\textwidth]{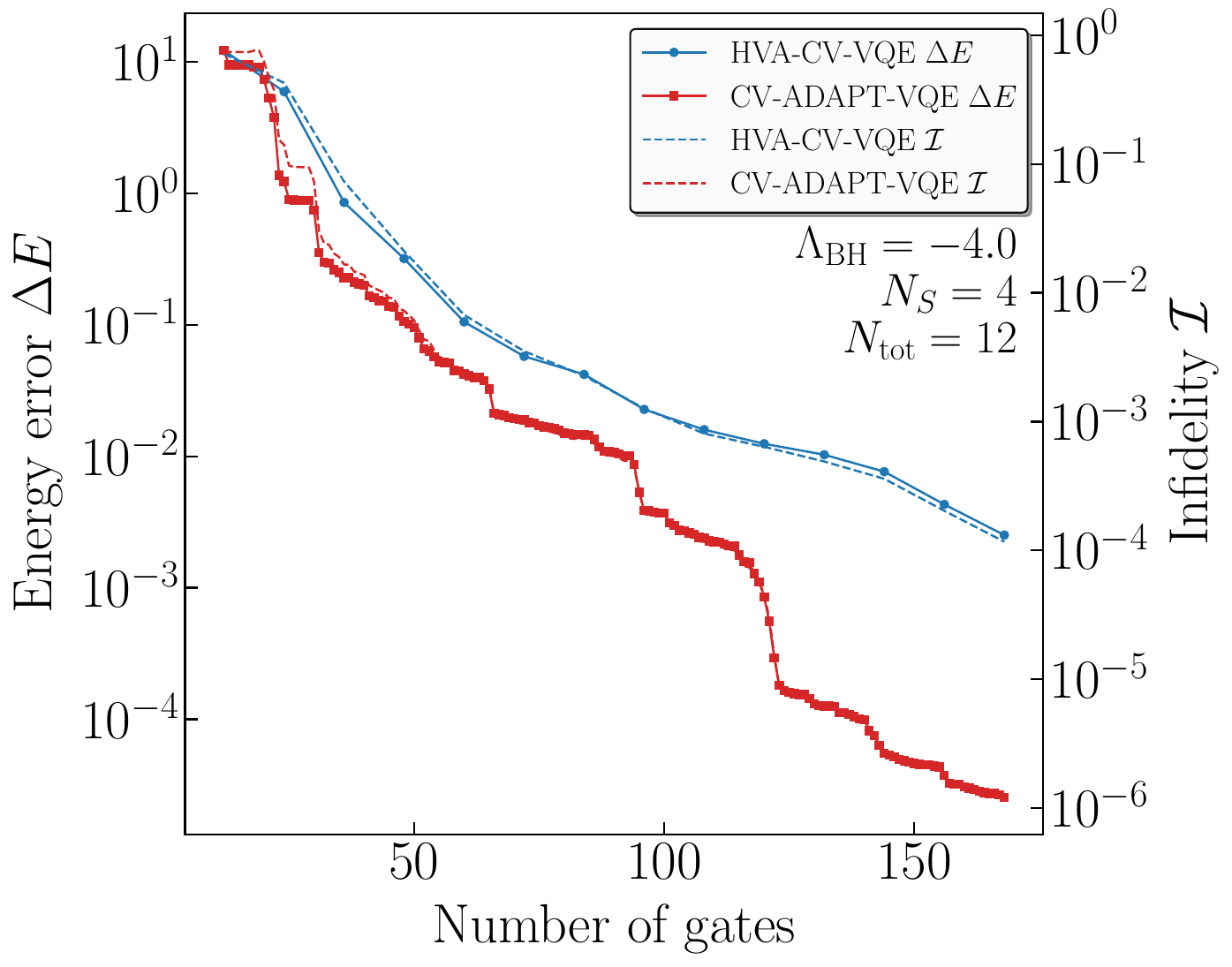}
        \label{fig:BH_REP}
    }
    \caption{\justifying The energy difference between the variational energy and the exact ground-state energy $\Delta E$ (solid lines, left axis), and the infidelity ${\cal I}$ (dashed lines, right axis), is shown as a function of the number of variational gates for the BH model. Panels (a) and (b) show the results for the attractive ($\Lambda_{\rm BH} = + 4.0$) and repulsive ($\Lambda_{\rm BH} = -4.0$) BH models, respectively, for $N_S = 4$ lattice sites and $N_{\text{tot}} = 12$ bosons with a local qumode cutoff of $D=13$. As an example, in panel (a), we also plot the magnitude of the gradient associated with the gate selected by ADAPT $|g_{r,{\rm init}}|$ (gray dotted line), i.e., the gate with the largest gradient at each iteration, using the scale of the left axis for reference.}

    \label{fig:BH_ENDIFF_VS_PARAMNUM}
\end{figure*}

\section{Numerical results~\label{sec:numerics}}

We start by describing the numerical setup of our classical simulations, and then present results for the ground-state preparation of the three bosonic lattice models considered in this work.

\subsection{Setup of the classical simulations~\label{sec:numerics_A}}

To perform the variational optimizations, we implement a custom differentiable simulator within \texttt{JAX}~\cite{jax2018github}. This choice enables execution on GPU hardware and allows for the efficient computation of exact energy gradients via automatic differentiation. 

In our implementation, we compute the unitary matrix elements using recurrence relations~\cite{Killoran_2019}, which is computationally more efficient than direct matrix exponentiation. These recursive definitions allow for the stable construction of operators with improved computational complexity. Both the quantum states and the Hamiltonian operators are represented as dense matrices in the Fock basis. While the physical Hamiltonians of the bosonic models considered here are inherently sparse, the current differentiable programming libraries lack support for sparse matrix operations. As a result, we employ a dense encoding where the memory size of a single operator scales as $\mathcal{O}(D^{2N_S})$, with $D$ denoting the local Fock cutoff and $N_S$ the number of sites. We execute our simulations on a NVIDIA A100 GPU, observing an empirical speedup of $\sim \mathcal{O}(10)$ for the largest system sizes compared to a CPU-based execution.

The dense representation imposes a strict memory bottleneck that restricts the accessible system size. For our maximal configuration of $N_S=5$ and $D=6$, the storage requirement is of order $1$ GB of VRAM per operator. Increasing the system size to $N_S=6$ would increase this requirement to the order of $10$ GB per operator, saturating the GPU memory bandwidth. This constraint restricts our exact numerical investigations to lattices of up to $N_S=5$. We note that a future transition to a fully sparse backend, where memory scaling is dictated by the state vector dimension $\mathcal{O}(D^{N_S})$ rather than the matrix dimension, would allow simulating significantly larger systems on the same hardware. We leave an exploration of this for future work.  

\begin{figure*}[t]
    \centering
        \subfigure{
        \includegraphics[width=0.99\textwidth]{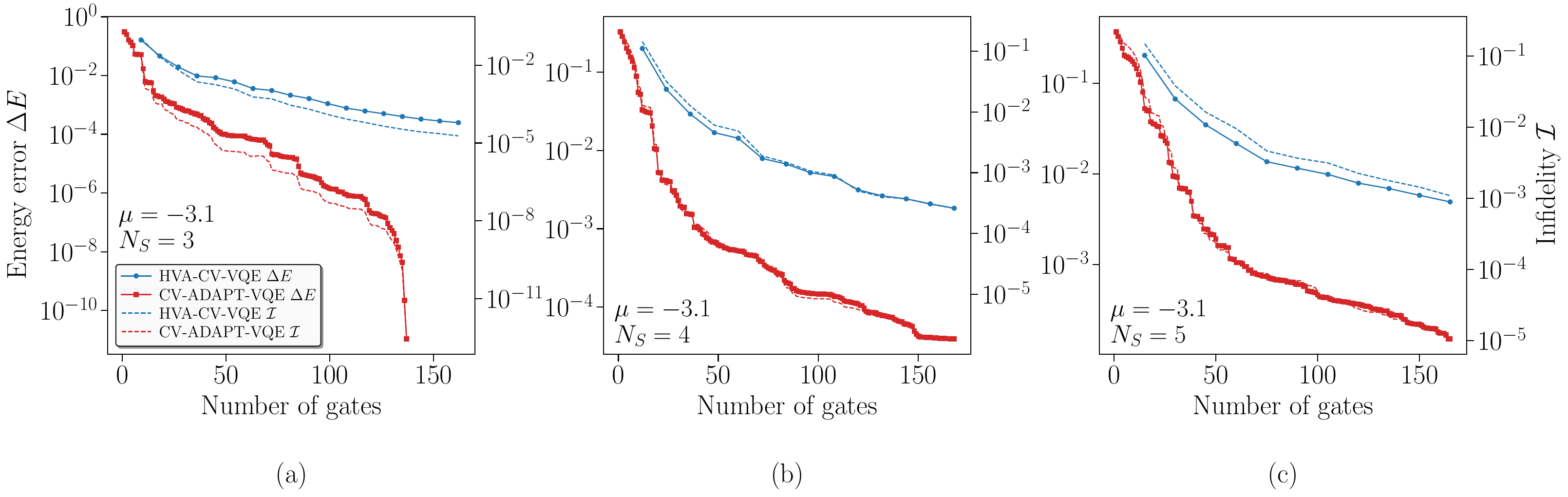}
        \label{fig:BKC_2}
    }

    \vspace{0.05cm} 

    \subfigure{
        \includegraphics[width=0.99\textwidth]{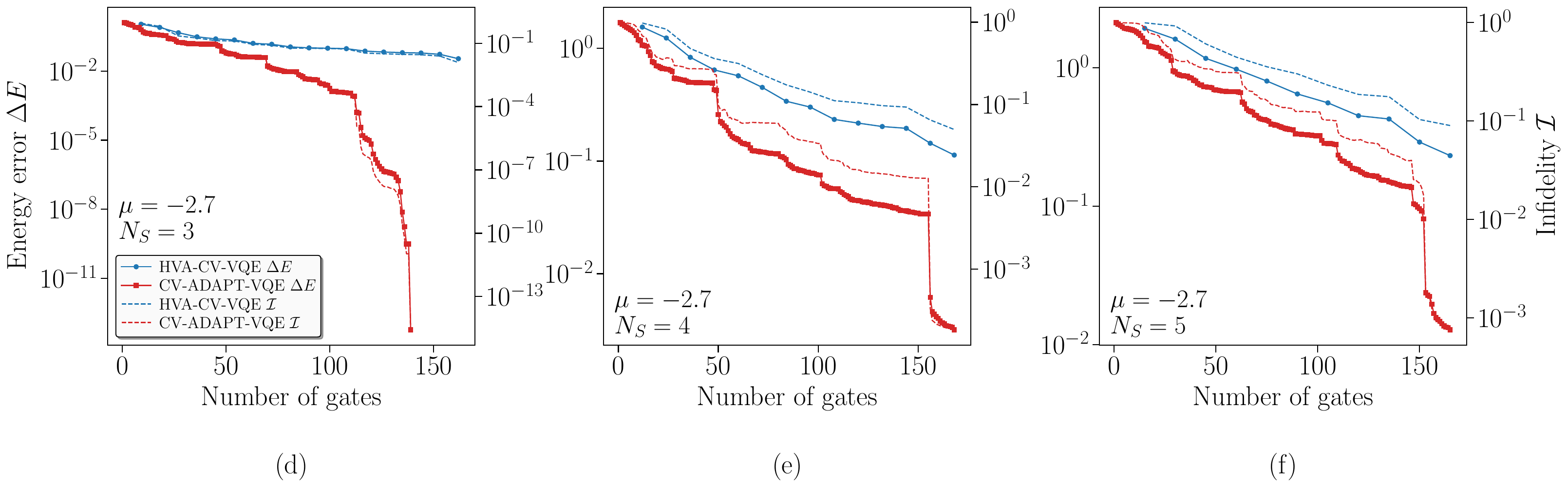}
        \label{fig:BKC_2}
    }
    \caption{\justifying
    The energy difference between the variational energy and the exact ground-state energy (solid lines, left axis), together with the infidelity (dashed lines, right axis), for the HVA-CV-VQE (blue) and CV-ADAPT-VQE (red) algorithms as a function of the number of gates for the BKC model. The model parameter choices for the two rows are $\mu = -3.1$ (top) and $\mu = -2.7$ (bottom), which are in the unstable and stable regimes, respectively, see Fig.~\ref{fig:bkc_ntot}. In each row, the number of lattice sites $N_S$ increases from 3 to 5 (left to right). For all simulations, the local qumode cutoff is $D = 6$.}
    \label{fig:BKC_ENDIFF_VS_PARAMNUM}
\end{figure*}

For the cost-function optimization, we employ the Broyden-Fletcher-Goldfarb-Shanno (BFGS) algorithm~\cite{virtanen_scipy_2020} to minimize the energy expectation value $E(\boldsymbol{\vartheta})$. For HVA-CV-VQE, we optimize each layer sequentially to improve performance \cite{skolik_2021}. Parameters for each newly added layer are initialized uniformly in the interval $[-0.05, 0.05]$, while previously optimized parameters are reused for the corresponding layers as the starting point. The optimization is terminated either after $1{,}000$ iterations or when the  $L^\infty$ norm of the gradient vector falls below $10^{-9}$. The choice of the random seed for this initialization of HVA-CV-VQE can affect the optimization performance and convergence.

For CV-ADAPT-VQE, the optimization for each fixed ansatz is terminated after either $1{,}000$ iterations or when the $L^\infty$ norm of the gradient vector falls below a variable tolerance $\min\!\left(10^{-6},\, |g_{r,{\rm init}}| \times 10^{-5}\right)$, where $g_{r,{\rm init}}$ denotes the gradient associated with the newly selected operator. This adaptive threshold is motivated by the exponential decay of gradients during the CV-ADAPT-VQE procedure. In particular, when the gradient associated with the newly selected operator is much smaller than the fixed BFGS stopping tolerance, the optimization can become insensitive to improvements along this direction and lead to premature plateaus in the CV-ADAPT-VQE procedure. As an example, we discuss the evolution of the gradient for the BH model below.

While the algorithm is insensitive to cutoff effects in the BH model, provided the cutoff is larger than the total excitation number, such effects are relevant for the two BKC models considered here. We note that the truncated squeezing operator preserves global parity even in the truncated Fock space, $[\hat P,\hat S_{\mathrm{trunc}}]=0$. However, the finite cutoff can affect the unitarity of the states when applying operators from the operator pool that change the occupation number. We address this issue by normalizing the state after each gate application.

We also attempted to enhance the expressivity of the operator pool by introducing two-gate blocks constructed from the single-gate operators already present in the pool. However, this modification did not lead to a noticeable improvement in the performance of CV-ADAPT-VQE.

\begin{figure*}[t]
    \centering
    \subfigure{
        \includegraphics[width=0.99\textwidth]{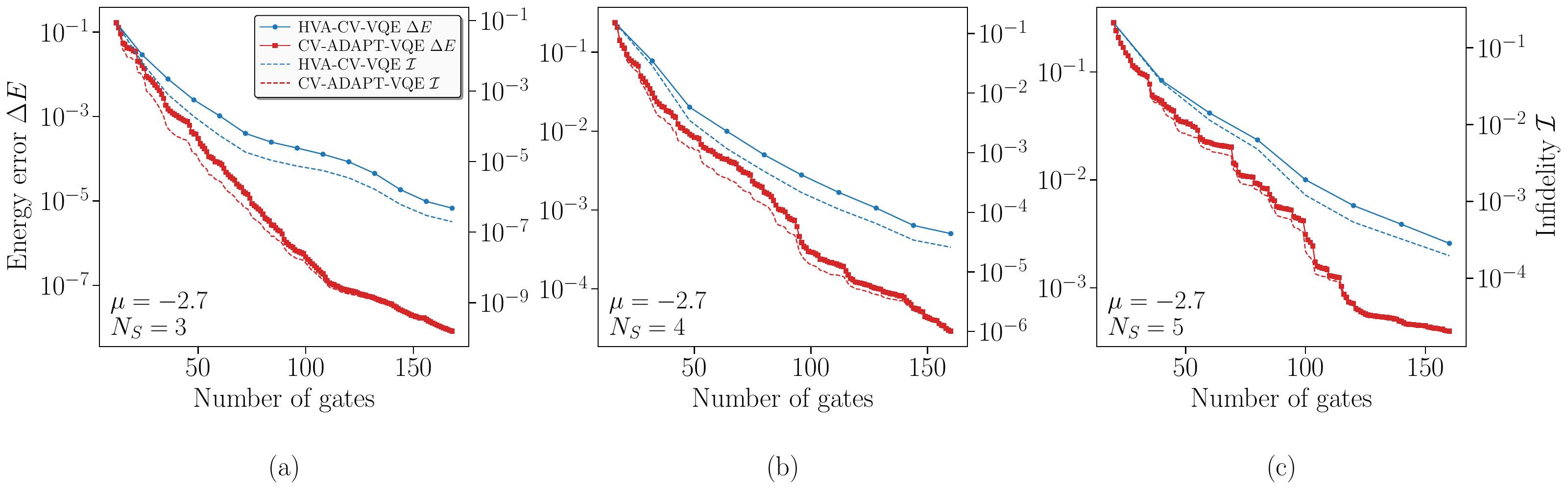}
         \label{fig:BKCint}
    }
    \caption{\justifying Difference between the variationally obtained energy and the result using exact diagonalization (solid lines, left axis), as well as the infidelity (dashed lines, right axis), for HVA-CV-VQE (blue) and CV-ADAPT-VQE (red) as a function of the number of gates for the BKC model with a repulsive on-site Kerr interaction $U=-1$. The results shown here correspond to the model parameters $(\mu=-2.7, \Delta=0.5)$, see the right panel in Fig.~\ref{fig:bkc_ntot}. From left to right, the number of lattice sites $N_S$ increases from 3 to 5.}
    \label{fig:BKCint_ENDIFF_VS_PARAMNUM}
\end{figure*}

Throughout this work, we initialize newly added variational ADAPT parameters to zero. We observe that initializing the parameters with small random values can improve convergence. However, choosing an initialization scale that is too large can lead to oscillatory behavior in the ground-state energy at later stages of the optimization. Since there is no clear a priori criterion for selecting an optimal initial scale, this choice would need to be determined by a hyperparameter search. We leave a more systematic study of the dependence on the initialization scale for future work.

\subsection{Bose-Hubbard model}

In Fig.~\ref{fig:BH_ENDIFF_VS_PARAMNUM}, we present the results for both the attractive ($\Lambda_{\rm BH} > 0$) and repulsive ($\Lambda_{\rm BH} < 0$) regimes of the BH model for $N_S=4$ lattice sites. The case $\Lambda_{\rm BH} = +4.0$ in Fig.~\ref{fig:BH_ATRAC} corresponds to a sector of the attractive regime where the ground state involves a relatively large number of bosonic configurations. For this particular case, we also plot $|g_{r,\mathrm{init}}|$, defined as the magnitude of the gradient associated with the gate selected by the ADAPT procedure at each iteration, in order to illustrate the behavior of the adaptive operator-selection process. The gradient generally decreases as the ansatz grows. Similar qualitative behavior is observed for the other $\Lambda_{\rm BH}$ values as well as for the BKC cases, and therefore, the gradient profile is only shown here as a representative example. In contrast, for $\Lambda_{\rm BH} = -4.0$, shown in Fig.~\ref{fig:BH_REP}, the strong repulsive interaction favors configurations with nearly uniform occupation across sites, resulting in a ground state dominated by only a small number of Fock configurations.

We quantify the performance of the two ground state preparation algorithms in terms of the energy difference between the variationally obtained result and the true ground state energy obtained using exact diagonalization $\Delta E=|E-E_0|$. In addition, we show the results for the infidelity $\mathcal{I} = 1 - \mathcal{F}$, where $\mathcal{F} = |\langle \psi_{0} | \psi_{\mathrm{VQE}} \rangle|^2$. Note that the loss function minimized for both algorithms is the energy. In both cases considered here, CV-ADAPT-VQE outperforms HVA-CV-VQE by roughly one to two orders of magnitude when the chosen maximum number of gates is reached. More generally, we observe that for both algorithms, the infidelity exhibits a dependence on the number of gates similar to that of the ground-state energy, even though it is not part of the objective function.

\subsection{Bosonic Kitaev chain}

We now turn to the numerical results for the BKC model. In Fig.~\ref{fig:BKC_ENDIFF_VS_PARAMNUM}, we consider two parameter choices near the stability boundary, one below $\left(\mu,\Delta\right) = (-3.1,0.5)$ and one above $\left(\mu,\Delta\right) = (-2.7,0.5)$, cf. Fig.~\ref{fig:bkc_ntot}. We compare the performance of the two algorithms as a function of the number of gates and the number of lattice sites $N_S$ for a local Hilbert space truncation of $D=6$.

For all lattice sizes considered here, CV-ADAPT-VQE consistently requires significantly fewer gates than HVA-CV-VQE to achieve the same level of accuracy. In particular, for a lattice with three sites ($N_S=3$), CV-ADAPT-VQE reaches machine precision after $\sim 140$ gates for both parameter choices. In contrast, HVA-CV-VQE plateaus at significantly larger values for both the energy difference and the infidelity. While the initial convergence of CV-ADAPT-VQE is slower for $\mu=-2.7$ compared to $\mu=-3.1$, the final convergence to approximate machine precision occurs after a comparable number of gates. This is consistent with the fact that the gaps of the two Hamiltonians are very similar. For larger lattices, we observe a similar pattern. The energy difference and infidelity for CV-ADAPT-VQE decrease more rapidly at early stages for $\mu=-3.1$. Also in this case, the gaps for both parameter choices are comparable, and achieving convergence to approximate machine precision would likely require additional gates and computational resources. Overall, the slower convergence observed for larger lattices is consistent with the larger Hilbert space and the increased complexity of the many-body ground state.

\subsection{Bosonic Kitaev chain with Kerr interaction}

Lastly, we consider the BKC model with an on-site Kerr interaction. Unlike the quadratic BKC Hamiltonian, the Kerr term renders the ground state non-Gaussian and therefore no longer amenable to an efficient classical Gaussian description. This makes the model a more challenging benchmark for variational state preparation within the truncated Hilbert space. In Fig.~\ref{fig:BKCint_ENDIFF_VS_PARAMNUM}, we compare the performance of HVA-CV-VQE and CV-ADAPT-VQE for a representative choice of parameters in the unstable regime of the Gaussian limit. Consistent with the non-interacting case, both the energy difference $\Delta E=|E - E_0|$ and the state infidelity $\mathcal{I} = 1 - \mathcal{F}$ are shown as a function of the number of gates.

Across all lattice sizes considered here, CV-ADAPT-VQE consistently outperforms HVA-CV-VQE, typically achieving a one- to two-order-of-magnitude improvement in both precision metrics. However, compared with the non-interacting case, the performance gap between the two algorithms is reduced, indicating that the inclusion of non-Gaussian interactions increases the complexity of optimal operator selection. Nevertheless, CV-ADAPT-VQE remains significantly more parameter-efficient, highlighting its robustness in the more challenging, interacting regime.

\section{Conclusions and outlook~\label{sec:conclusions}}

In this work, we presented CV-ADAPT-VQE, an adaptive variational quantum algorithm for continuous-variable quantum computing, and demonstrated its effectiveness for ground-state preparation in bosonic lattice models. By constructing symmetry-preserving operator pools for the Bose-Hubbard and bosonic Kitaev chain models, we showed that the adaptive approach consistently outperforms HVA-CV-VQE in both accuracy and circuit depth. In particular, CV-ADAPT-VQE achieves significantly lower energy errors and infidelities with fewer gates, including in regimes with strong correlations, finite local Hilbert-space truncations, and non-Gaussian features induced by Kerr interactions. These results highlight the advantage of adaptive ansätze in continuous-variable quantum computing, where large Hilbert spaces and non-Gaussian interactions can make fixed circuit structures inefficient. More broadly, they demonstrate the importance of tailoring symmetry-preserving operator pools to the structure of the target model for efficient ground-state preparation.

More broadly, this work opens several promising directions at the interface of algorithm design, numerical simulation, and hardware implementation. On the algorithmic side, resource-aware and co-design extensions of ADAPT-VQE~\cite{PhysRevResearch.6.013254,ramoa_co-designed_2026} offer a pathway to reduce measurement costs while accounting for hardware constraints. We expect this to be particularly relevant in a hybrid qubit-qumode setting. We also find that design choices such as the initialization of newly added parameters, set to zero in our implementation, can influence convergence, suggesting that more systematic and potentially adaptive strategies may further improve performance. On the computational side, our current implementation is limited by dense representations, motivating the development of more efficient sparse encodings and the integration with scalable ADAPT-VQE approaches~\cite{Farrell:2023fgd,gustafson_surrogate_2024} to access larger system sizes and enable extrapolations to physically relevant regimes.

At the same time, these results point toward direct opportunities for simulations on near-term quantum hardware, where significant recent progress has been made in qumode-based quantum simulations across leading platforms. Adapting the operator pools to hardware-native gate sets, incorporating experimental constraints and noise directly into the ansatz construction and gate selection, and extending the framework to hybrid qubit-qumode architectures all provide natural next steps. Together, these directions suggest that CV-ADAPT-VQE can serve as a useful and flexible tool for applications in condensed matter physics, quantum chemistry, and quantum field theory.

\section*{Code availability}

The code developed for this work is available on GitHub at: \url{https://github.com/gloriatejegar/CV_ADAPT_VQE}.

\section*{Acknowledgements}

We would like to thank Abhijit Chakraborty, Tommaso Rainaldi, and Karunya Shirali for helpful discussions. The research was supported by the DOE, Office of Science, Office of Nuclear Physics, Early Career Program under contract DE-SC0025881 and by the Center for Distributed Quantum Processing at Stony Brook University. This research used resources of the National Energy Research Scientific Computing Center (NERSC), a DOE Office of Science User Facility using NERSC award NP-ERCAP0037787. The authors would like to thank Stony Brook Research Computing and Cyberinfrastructure, and the Institute for Advanced Computational Science at Stony Brook University, for access to the SeaWulf computing system, made possible by grants from the National Science Foundation (\#1531492 and Major Research Instrumentation award \#2215987), with matching funds from Empire State Development’s Division of Science, Technology and Innovation (NYSTAR) program (contract C210148). J. Y. Araz is supported by the Institute for Particle Physics Phenomenology Associateship Scheme and by the Royal Society under grant no. IES/R2/252139. M. Ram\^{o}a, B. Sambasivam, and S. E. Economou are supported by the U.S. Department of Energy, Office of Science, Office of Advanced Scientific Computing Research under Award Number DE-SC0025430.

\section*{Disclaimer} 
This paper was prepared for information purposes and is not a product of HSBC Bank Plc. or its affiliates. Neither HSBC Bank Plc. nor any of its affiliates make any explicit or implied representation or warranty, and none of them accept any liability in connection with this paper, including, but not limited to, the completeness, accuracy, reliability of information contained herein, and the potential legal and compliance effects thereof.

\appendix

\section{Stability condition of the Bosonic Kitaev chain~\label{app1}}

We work with periodic boundary conditions by identifying the bosonic ladder operators $\hat a_{N_S+1} = \hat a_{1}$. A similar derivation was presented in Ref.~\cite{wang_quantum_2022} for $\mu=0$. Using the Fourier transformation
\begin{align}
    \hat a_r = \frac{1}{\sqrt{N_S}} \sum_{k} e^{ikr} \hat a_k\,,
\end{align}
where $k=\frac{2 \pi}{N_S} n$  and $n=0,1, \ldots, N_S-1$. We can rewrite the Hamiltonian above in terms of the Bogoliubov-de Gennes (BdG) form as
\begin{align}
    \hat H_{\rm BKC} = \frac{1}{2} \sum_{k} 
    \begin{pmatrix}
        \hat a_k^\dagger & \hat a_{-k}
    \end{pmatrix}
    H_{\text{BdG}}(k)
    \begin{pmatrix}
        \hat a_k \\
        \hat a_{-k}^\dagger
    \end{pmatrix}
    - \text{Tr}(\mu \mathbb{I})\,,
\end{align}
where
\begin{align}\label{eq:BdGHamiltonian}
    H_{\text{BdG}}(k) =&\, - (2J \cos k + \mu) \sigma_0 - 2 \Delta \cos k \sigma_x\,.
\end{align}
Here, $\mathbb{I}$ denotes the identity and $\sigma_{x,y,z}$ are the Pauli matrices. The generalized eigenvalue problem can be written as
\begin{align}\label{eq:gen_eig}
    H_{\text{BdG}}(k)\, v = E\, \sigma_z v\,,
\end{align}
which allows us to solve for the energy spectrum of the bosonic Kitaev chain using the characteristic equation
\begin{align}\label{eq:char_eq}
    \det\!\big[\sigma_z H_{\text{BdG}}(k) - E(k)\, \mathbb{I}\big] = 0\,.
\end{align}
This leads to the result referred to in the main text
\begin{align}\label{eq:dispersion}
&\,E(k) =  \pm \sqrt{\big(-2J \cos k - \mu\big)^2 
       - \big(2\Delta \cos k \big)^2 }\,.
\end{align}

\section{Operator tiling~\label{app2}}

In this appendix, we implement operator tiling, originally introduced for qubit-based lattice models in Ref.~\cite{VanDykePRRTiling}. In general, operator tiling applies to locally interacting translationally invariant systems. The approach uses a classical preprocessing step to learn operators that are important in preparing the ground state. The constructed operator pool is guaranteed to be expressive enough to reach the ground state and to scale efficiently with the system size, $N$, as $\mathcal{O}(N)$. Operator tiling involves the following steps:
\begin{enumerate}
    \item Run CV-ADAPT-VQE for a classically tractable simulation of the system on a small, non-trivial lattice of size $N_{\text{small}}$ using an expressive operator pool.
    \item Collect the operators that ADAPT chooses.
    \item Repeat step 1-2 until the different operators with degenerate gradients are collected.
    \item Tile these ``small'' operators along the lattice, padding with identity operators to match the size of the larger, target system, $N_{\text{large}}$. This forms the operator pool for the larger system. 
\end{enumerate}

\begin{figure}
    \centering
    \includegraphics[width=\linewidth]{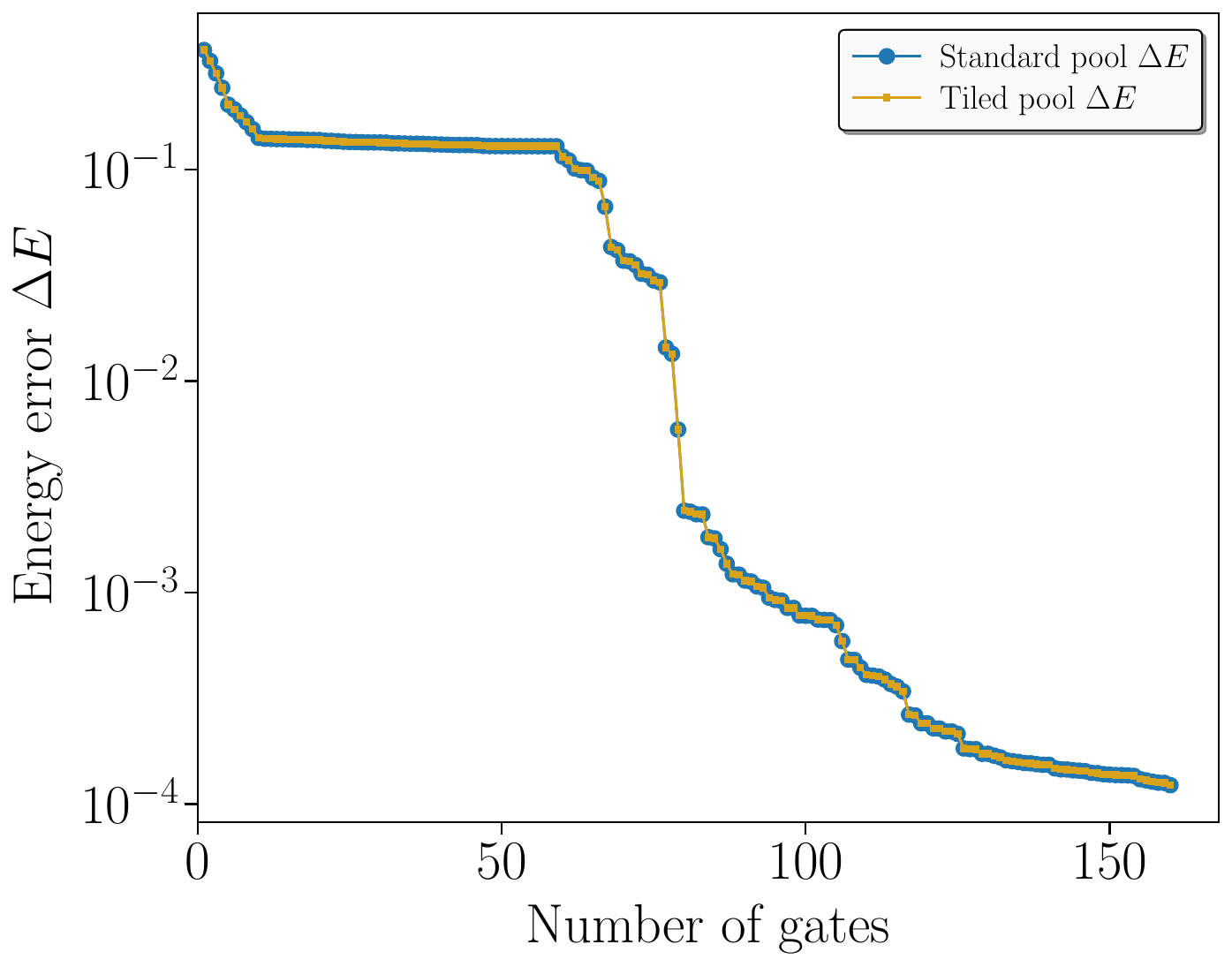}
    \caption{\justifying Energy difference with respect to exact diagonalization (left axis) and infidelity (right axis) as a function of the number of gates in the ansatz for the BKC model with $N_S=5, \Delta=0.5, \mu=-3.1, D=6$. The comparison is between the regular full pool used in the main text (with size 30) and the operator tiling pool (with size 25).}
    \label{fig:TilingBKC}
\end{figure}

In Fig.~\ref{fig:TilingBKC}, we show a comparison between using this tiled pool and the regular pool used in the main text for the BKC model with $N_S=5, \Delta=0.5, \mu=-3.1, D=6$. The operators chosen and the subsequent convergence of the two approaches are fully identical to each other. The only difference is the size of the two pools, as shown in the legend of Fig.~\ref{fig:TilingBKC}. The experiments on smaller system sizes successfully identified operators that are unimportant to the system, making the size of the tiled pool smaller. This will lead to savings in the number of gradients that need to be measured at each adaptive iteration. In all the results presented in this work, the full operator pool scales linearly with system size, so tiling amounts to a constant saving in the asymptotic sense. However, for models where a more expressive local pool is necessary, we expect the gains from operator tiling to be more pronounced. 

\begin{figure*}[t]
    \subfigure[]{
        \includegraphics[width=0.4\textwidth]{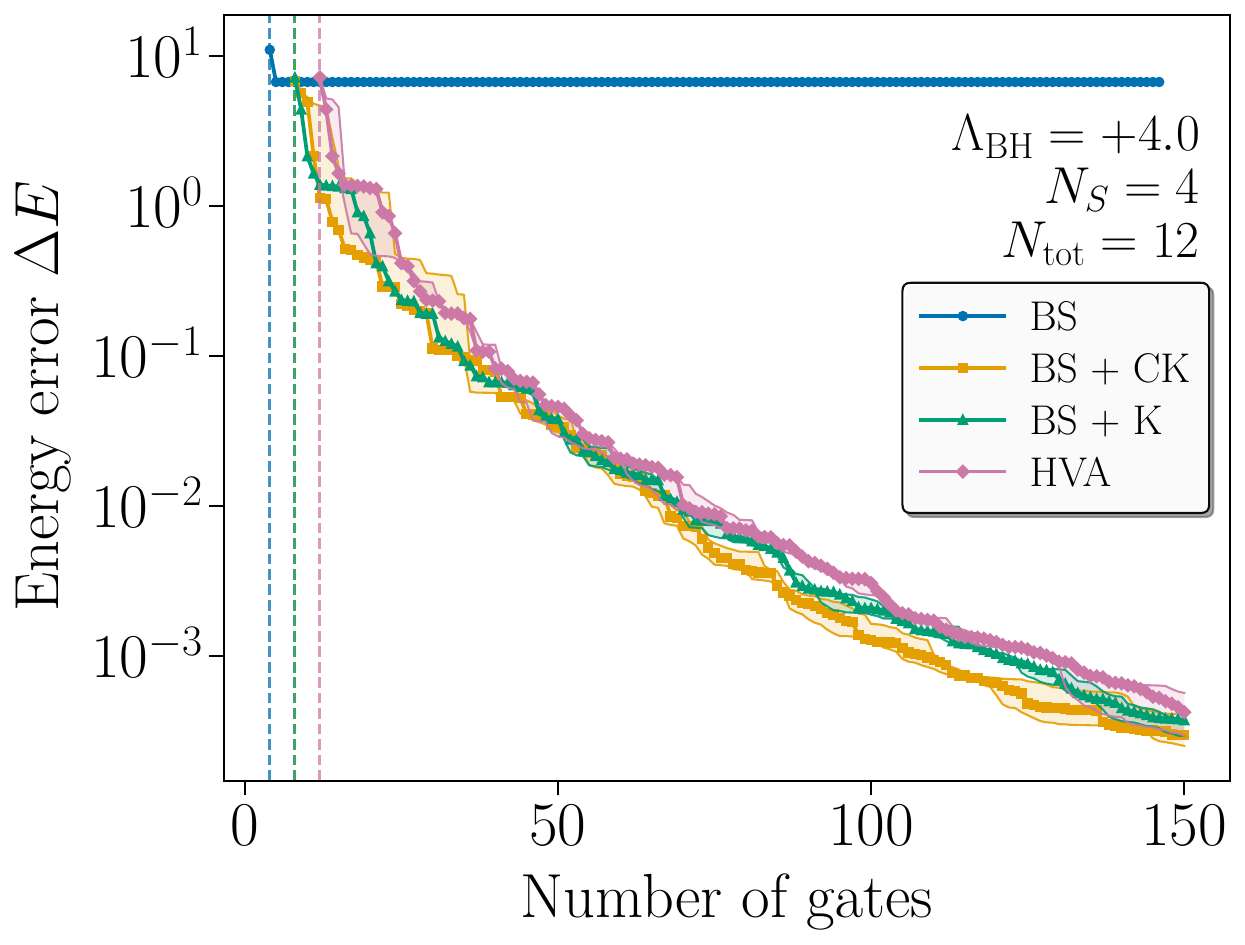}
        \label{fig:BH_ATRAC_warmup}}
    \subfigure[]{
        \includegraphics[width=0.4\textwidth]{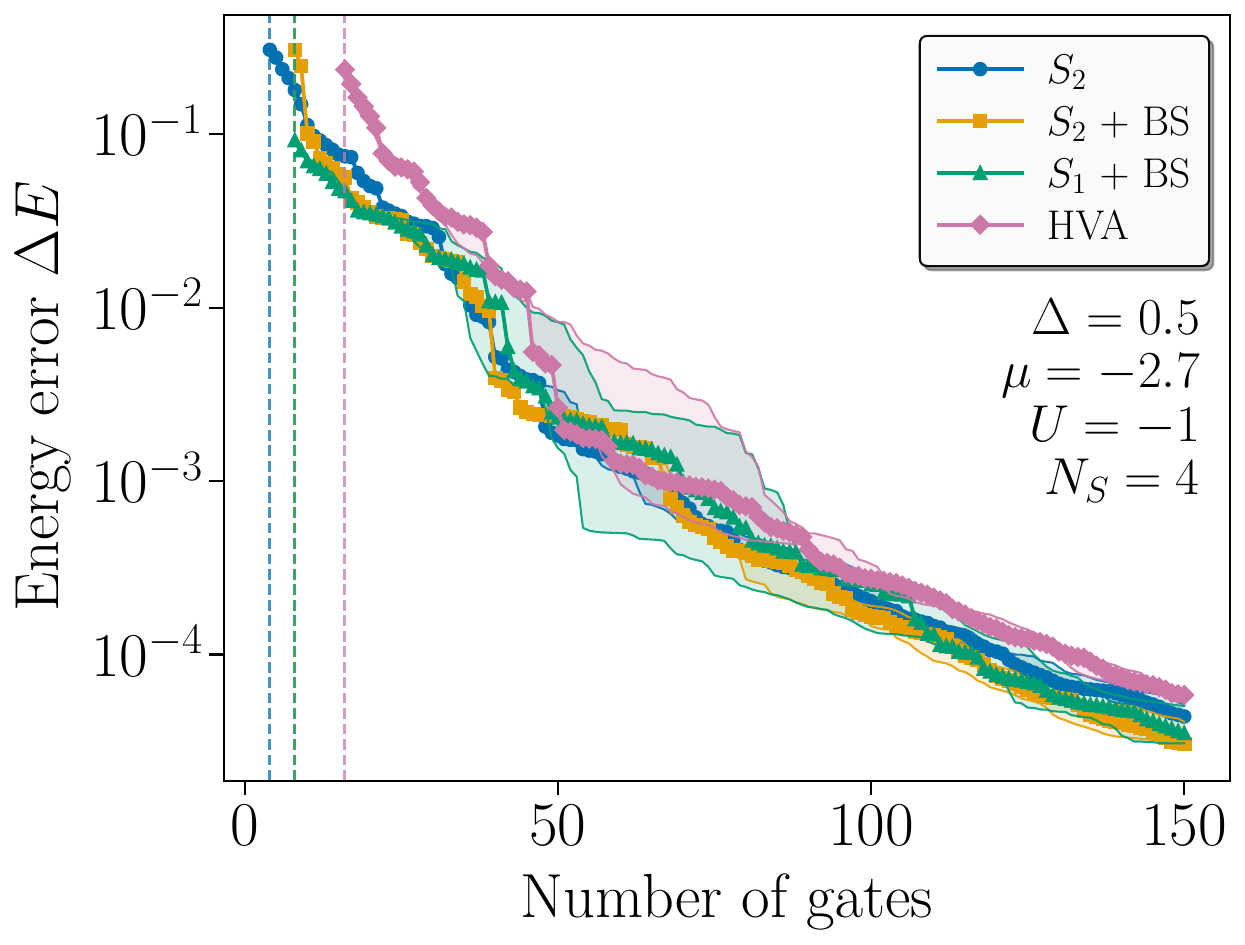}
        \label{fig:BKCint_warmup}}
\caption{\justifying The energy difference between the variational energy and the exact ground-state energy, $\Delta E$, for CV-ADAPT-VQE using different warm-up circuits. Panels (a) and (b) show results for the attractive BH model ($\Lambda_{\rm BH}=+4.0$) and the BKC+Kerr model, respectively, both with $N_S=4$ lattice sites. Each configuration was evaluated using three different random seeds. The solid curves denote the median trajectory, while the shaded regions span the minimum and maximum trajectories across the three runs.}
\label{fig:warm_up_choices}
\end{figure*}

\begin{figure*}[t]
    \subfigure[]{
        \includegraphics[width=0.4\textwidth]{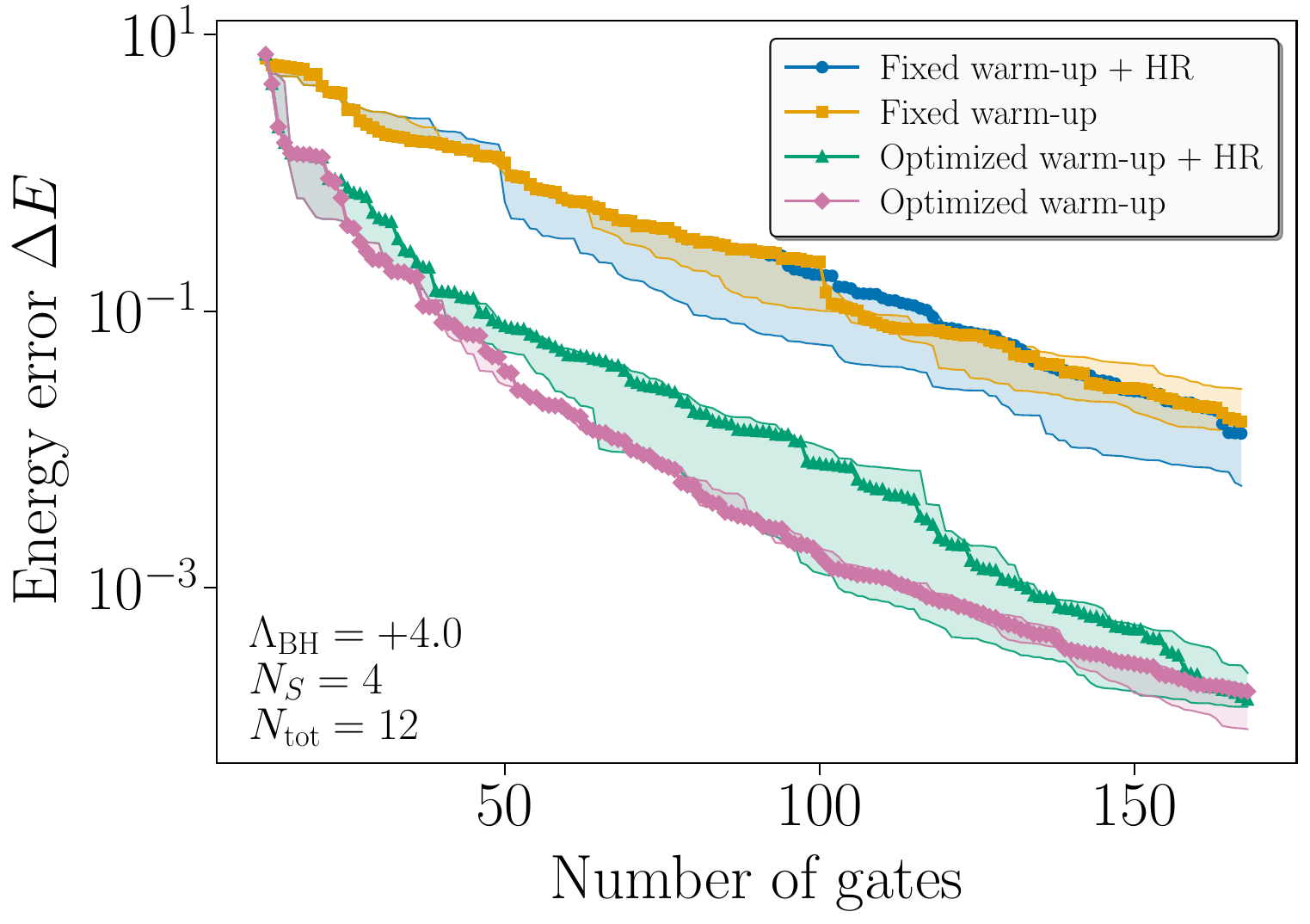}
        \label{fig:BH_RH_VS_noRH}}
    \vspace{0.01mm}
    \subfigure[]{
        \includegraphics[width=0.4\textwidth]{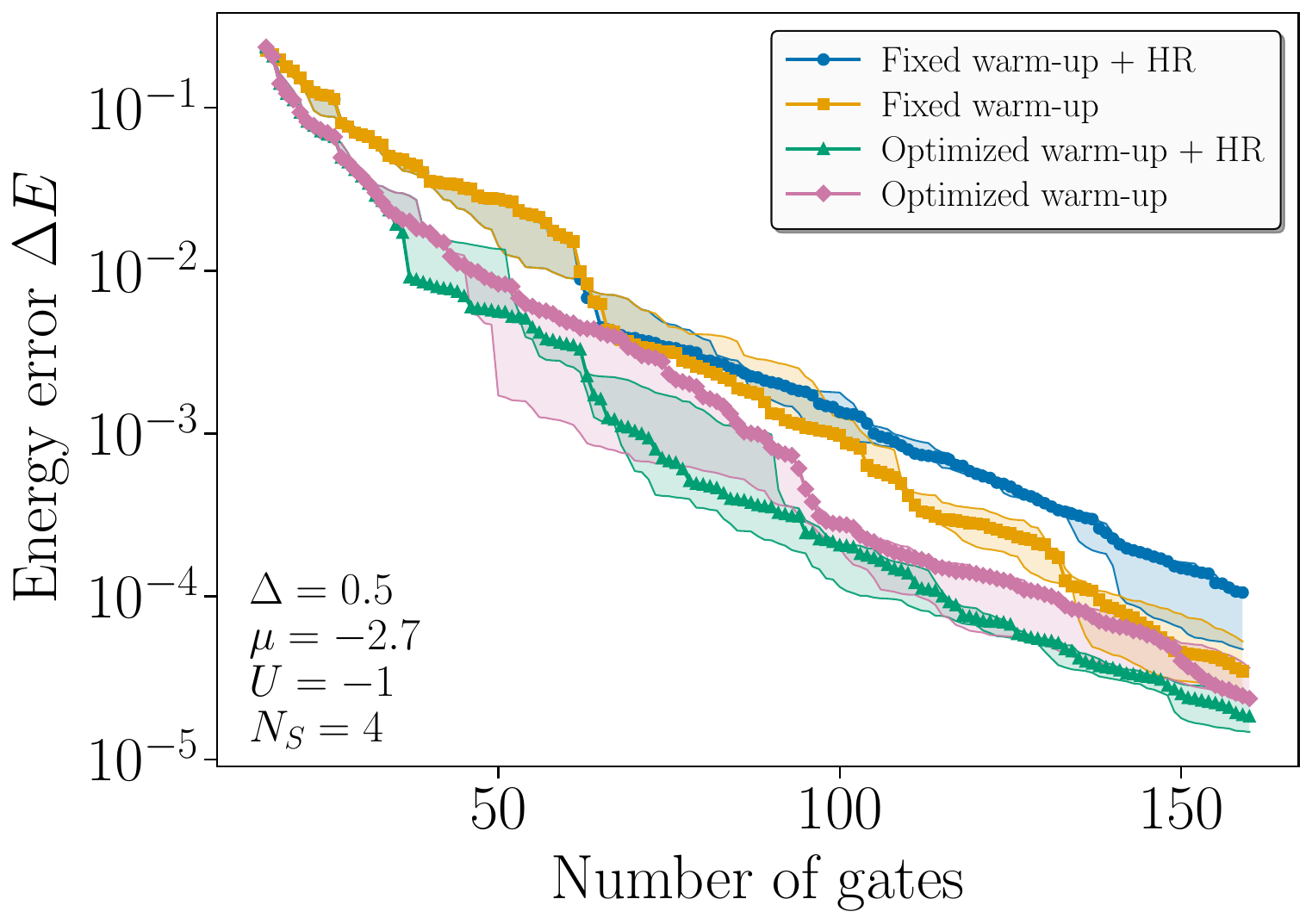}
        \label{fig:BKCint_RH_VS_noRH}}
    \caption{\justifying The energy difference between the variational energy and the exact ground-state energy, $\Delta E$, as a function of the number of gates for CV-ADAPT-VQE using different optimization strategies. The comparison includes four cases: fixed warm-up parameters with Hessian recycling (HR), fixed warm-up parameters without Hessian recycling, optimized warm-up parameters within ADAPT with Hessian recycling, and optimized warm-up parameters within ADAPT without Hessian recycling. The solid curves correspond to the pointwise median over different random seeds, while the shaded regions indicate the minimum and maximum values across the seeds at each gate point. Panel (a) shows the results for the attractive BH model ($\Lambda_{\rm BH}=+4.0$), while panel (b) corresponds to the BKC model with on-site interaction, both for $N_S=4$ lattice sites.
    }
    \label{fig:RH_VS_noRH}
\end{figure*}

\section{Choice of the warm-up circuit~\label{app:warmup}}

\begin{figure*}[t]
    \subfigure[]{
        \includegraphics[width=0.4\textwidth]{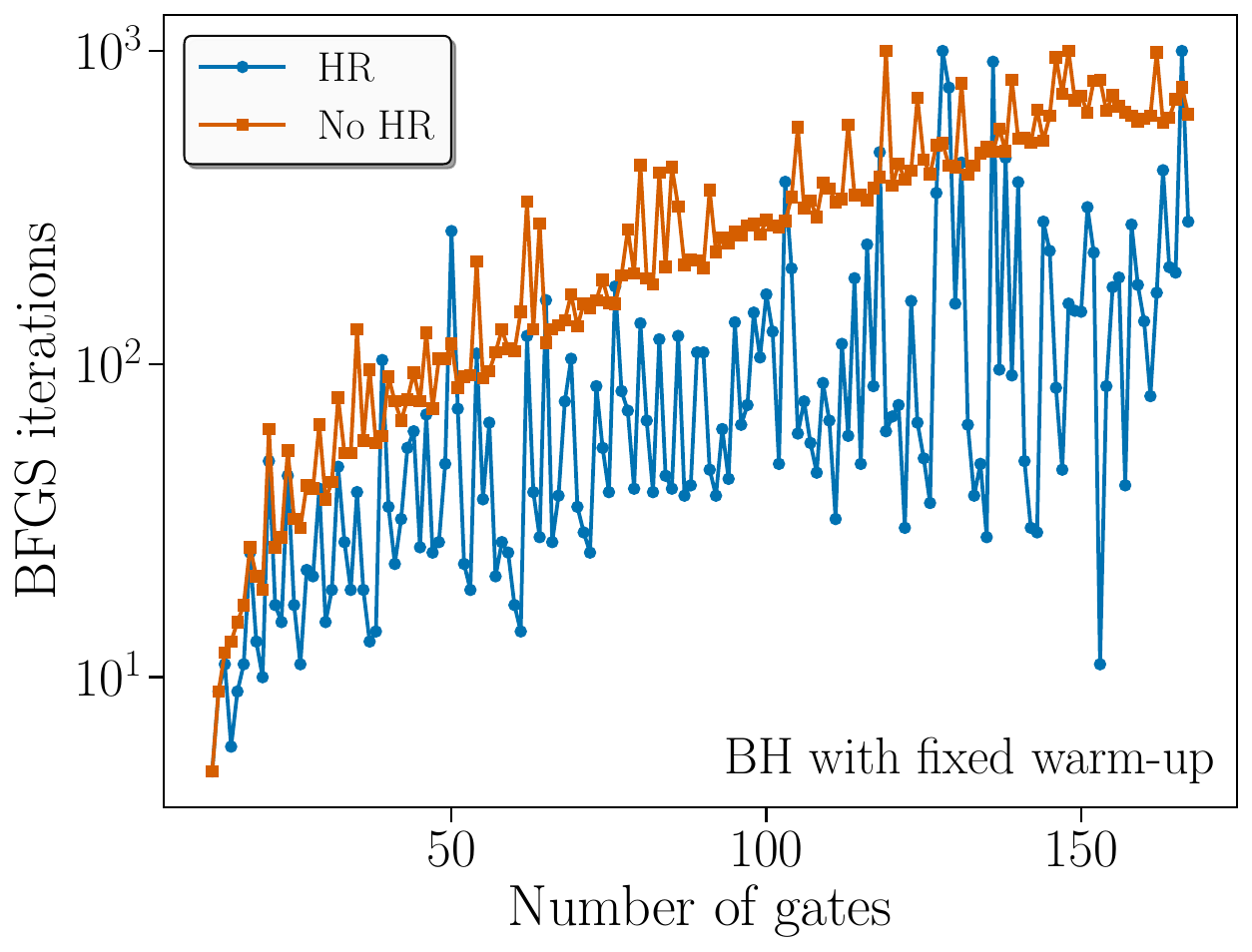}
        \label{fig:BH_RH_iter }}
    \subfigure[]{
        \includegraphics[width=0.4\textwidth]{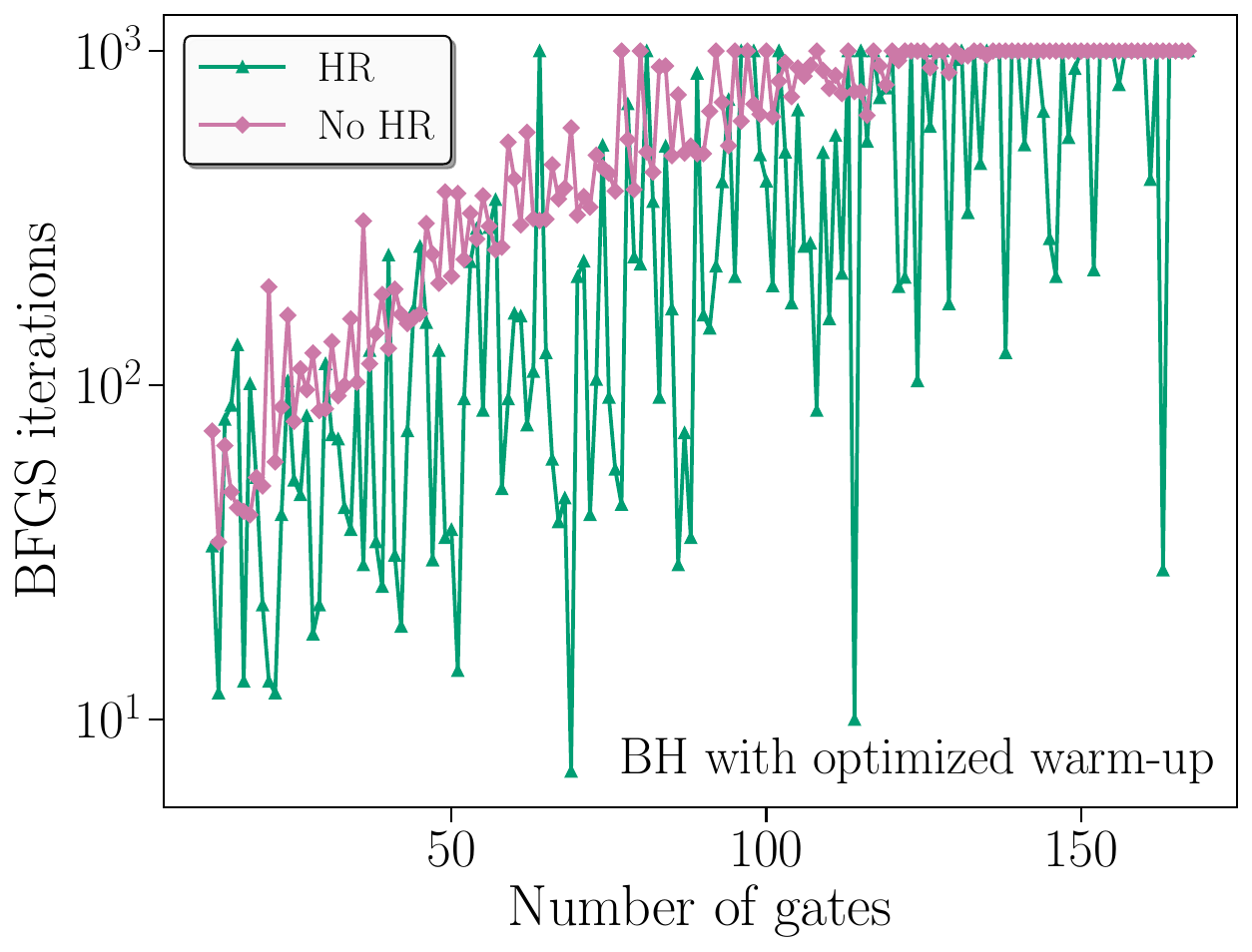}
        \label{fig:BH_noRH_iter}}

    \vspace{0.5cm}
    \subfigure[]{
        \includegraphics[width=0.4\textwidth]{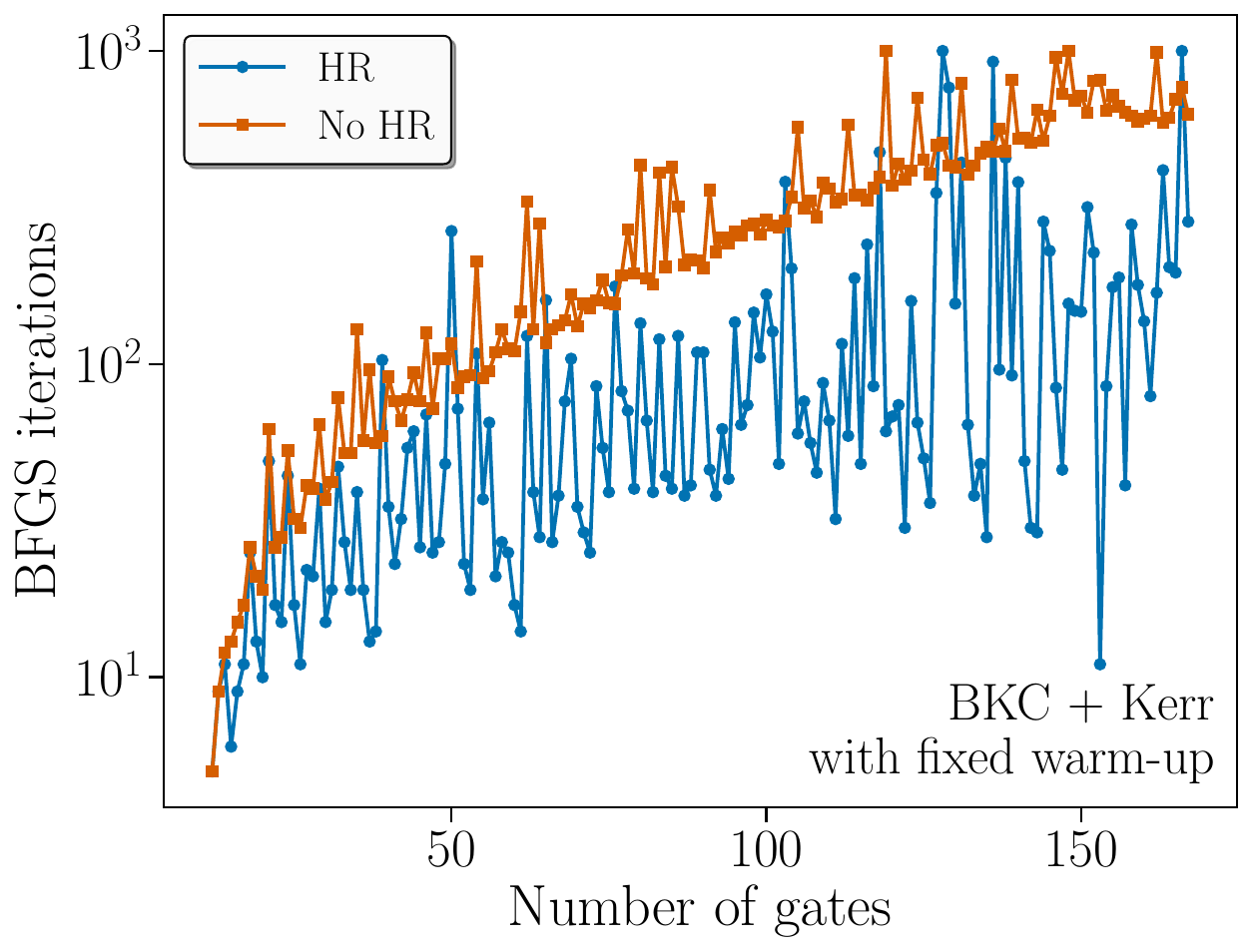}
        \label{fig:BKCint_RH_iter }}
    \subfigure[]{
        \includegraphics[width=0.4\textwidth]{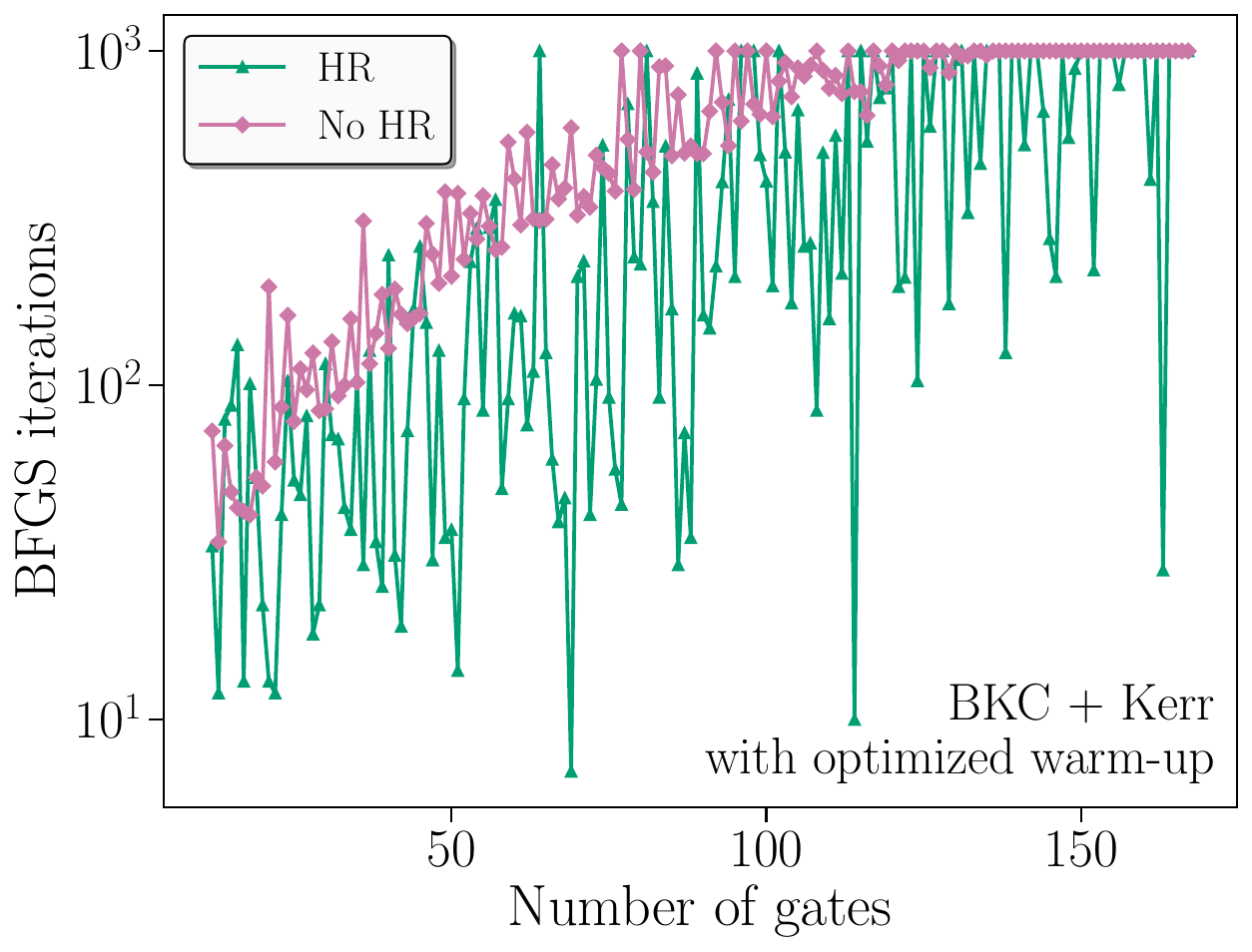}
        \label{fig:BKCint_noRH_iter}}

\caption{\justifying BFGS iterations as a function of the number of gates for CV-ADAPT-VQE with and without Hessian recycling (HR). Panels (a) and (b) correspond to the attractive BH model ($\Lambda_{\rm BH}=+4.0$) with fixed and optimized warm-up parameters within ADAPT, respectively. Panels (c) and (d) show the corresponding results for the BKC model with on-site interaction. In all cases, the warm-up layer is given by an HVA circuit, and $N_S=4$ lattice sites are used. For each configuration, the seed with the best performance was selected, and the maximum number of BFGS iterations was set to 1000.
}
\label{fig:RH_VS_noRH_iter}
\end{figure*}

As discussed in the main text, the BH and BKC with an on-site interaction models require a warm-up circuit to initialize the CV-ADAPT-VQE optimization, for which we employed a single HVA layer. To test whether this choice significantly affects the final result, we compare to four physics-motivated warm-up circuits for each model. For the BH model, we consider the following warm-up circuits: a single HVA layer, a ring beam splitter ($\mathrm{BS}$), the beam splitter ring followed by an on-site Kerr layer ($\mathrm{BS}+\mathrm{K}$), and the beam splitter ring followed by a nearest-neighbor cross-Kerr ring ($\mathrm{BS}+\mathrm{CK}$). For the BKC model with on-site interaction, we consider: a single HVA layer, a two-mode squeezing ring ($\mathrm{S}_2$), the $\mathrm{S}_2$ ring followed by a beam splitter ring ($\mathrm{S}_2+\mathrm{BS}$), and a single-mode squeezing layer applied on every site followed by a beam splitter ring ($\mathrm{S}_1+\mathrm{BS}$). All circuit parameters were initialized uniformly in $[-0.05, 0.05]$, and the BFGS optimization at each ADAPT step was terminated after $500$ iterations or once the $L^{\infty}$ gradient norm falls below $10^{-9}$. The results are shown in Fig.~\ref{fig:warm_up_choices}.

Although the initial energy errors differ by up to roughly half an order of magnitude across the various warm-up circuits, the corresponding trajectories ultimately converge to comparable final values of $\Delta E$ after a sufficient number of ADAPT iterations. The only exception is the beam splitter ring warm start for the BH model, where the optimization gets trapped in an excited state. These results suggest that, for small lattice sizes, CV-ADAPT-VQE is sufficiently expressive to recover from suboptimal warm-up choices, provided that the initial warm-start ansatz captures the correct qualitative structure of the target state.

\section{Hessian recycling~\label{app:HessianRecycling}}

Reference~\cite{ramoa_reducing_2025} introduced a Hessian recycling technique that modifies quasi-Newton optimization protocols to allow information about second-order derivatives of the cost function to flow from one ADAPT-VQE iteration to the next. For molecular simulations, this was shown to lead to a significantly lower error than the canonical ADAPT-VQE algorithm while resulting in the same ansatz and energy accuracy. In this appendix, we showcase the impact of this technique in CV-ADAPT-VQE.

It is important to note that for two of the models considered in this paper (BH and BKC with on-site interactions), our adaptive algorithm includes a warm-up circuit consisting of a layer of the HVA ansatz, with the corresponding parameters being reoptimized in each ADAPT-VQE iteration (together with the newly added variational parameters). This is in contrast to the original version of the ADAPT-VQE algorithm, where a parameter-free reference state is used as the starting point for the adaptive ansatz construction (representing, in the case of electronic structure problems, the mean-field approximation to the ground state); in this case, the initial state remains unchanged throughout the algorithm, unlike what happens with our warm-up HVA layer. Cost landscapes associated with the HVA typically have many local minima, and the parameters may vary significantly from one optimization to the next; hence, it is possible that constantly re-optimizing these parameters will lead to a local cost landscape with significantly different features, jeopardizing the effectiveness of Hessian recycling. Taking this into consideration, we consider two versions of CV-ADAPT-VQE: one where the warm-up circuit is continuously optimized (as in the main text), and one where it is fixed after the initial optimization.

Fig.~\ref{fig:RH_VS_noRH} shows the CV-ADAPT-VQE error convergence plots with fixed/non-fixed warm-up parameters and with/without Hessian recycling, for the two models that use the HVA warm-up layer (BCH with Kerr interaction and BH). We observe that the error associated with the continuous re-optimization of the warm-up layer is lower; this is to be expected, since the variational space associated with the fixed-layer method corresponds to a restricted subspace of the one reachable with the non-fixed warm-up layer parameters. Additionally, we can see that the errors with and without Hessian recycling diverge after a given number of iterations in all cases. This is in contrast with the case of molecular simulations, where the application of this technique does not typically affect accuracy. However, we note that it is very easy for this to happen, especially with finite-precision expectation values; a minor shift in candidate gradients may result in a different operator being selected, after which point the constructed ansatz may be entirely changed. We have no reason to believe that Hessian recycling would improve or worsen the quality of the minima found by the optimizer at each step; on average, we expect that the two methods will lead to a similar error. This is confirmed by numerical simulations, where Hessian recycling leads to lower or higher errors depending on the particular system and trial.

The purpose of Hessian recycling is not to improve accuracy, but to decrease measurement costs; hence, in Fig.~\ref{fig:RH_VS_noRH_iter}, we compare the number of optimizer iterations with and without Hessian recycling in the same settings as before. In each case, we try multiple random seeds and select the one leading to the best result. We observe that in all cases, Hessian recycling significantly reduces the number of optimizer iterations per ADAPT-VQE iteration, which in practice will translate to a decrease in the number of calls to the quantum computer, as desired.

\bibliographystyle{utphys}
\bibliography{bibliography}

\end{document}